\documentclass[11pt]{article}
\usepackage{color, amsmath, amssymb, amsbsy, amsthm, graphicx, bbm, amsfonts, bm, dsfont}
\usepackage{wrapfig}
\usepackage{setspace, booktabs, comment, graphicx, algorithm,  latexsym}
\usepackage[top=1in, bottom=1in, left=1in, right=1in]{geometry}
\usepackage[toc,page]{appendix}
\RequirePackage[numbers]{natbib}
\RequirePackage[colorlinks,allcolors=blue,urlcolor=blue]{hyperref}
\usepackage{graphicx}
\usepackage{caption} 
\usepackage{wrapfig}
\usepackage{enumitem}

\usepackage{xr}
\makeatletter

\newcommand*{\addFileDependency}[1]{
\typeout{(#1)}
\@addtofilelist{#1}
\IfFileExists{#1}{}{\typeout{No file #1.}}
}\makeatother

\usepackage{algpseudocode}
\usepackage{diagbox}
\usepackage{adjustbox}
\usepackage{wrapfig}
\usepackage{lipsum}
\usepackage{algpseudocode}
\usepackage{algorithm}
\usepackage{multirow}

\allowdisplaybreaks

\theoremstyle{plain}
\newtheorem{theorem}{Theorem}
\newtheorem{corollary}{Corollary}
\newtheorem{remark}{Remark}
\newtheorem{assumption}{Assumption}
 
\newtheorem{lemma}{Lemma}

\newtheorem{proposition}{Proposition}
\newtheorem*{stage1}{Stage $\bm{1}$}
\newtheorem*{staget}{Stage $\bm{t}$, for $\bm{t=2,\ldots, T-1}$}
\newtheorem*{stageT}{Statistical inference after Stage $\bm{T}$}

\def \hat{\widehat}
\def \tilde{\widetilde}
\def \bbp{ \mathbb{P}}

\def \bbe{\mathbb{E}}

\def\bbe{\mathbb{E}}
\def\bbv{\mathbb{V}}

\DeclareMathOperator*{\argmax}{argmax}
\usepackage{tikz}
\usetikzlibrary{shapes,decorations,arrows,calc,arrows.meta,fit,positioning}
\tikzset{
	-Latex,auto,node distance =1 cm and 1 cm,semithick,
	state/.style ={ellipse, draw, minimum width = 0.7 cm},
	point/.style = {circle, draw, inner sep=0.04cm,fill,node contents={}},
	bidirected/.style={Latex-Latex,dashed},
	el/.style = {inner sep=2pt, align=left, sloped}
}

\newcommand*{\refeq}[1]{%
  \begingroup
    \hypersetup{
      linkcolor=black,
      linkbordercolor=black,
    }%
    \ref{#1}%
  \endgroup
}

\def\Variance{\mathbb{V}}

\makeatletter
\newcommand{\leqnomode}{\tagsleft@true}
\newcommand{\reqnomode}{\tagsleft@false}
\makeatother

\begin{document}
	\title{Adaptive Experiments Toward Learning Treatment Effect Heterogeneity}
	\author{Waverly Wei$^1$ \and Xinwei Ma$^2$ \and Jingshen Wang$^1$\thanks{Correspondence: jingshenwang@berkeley.edu.}}
	\date{ $^1$Division of Biostatistics, University of California, Berkeley, California, United States\\
 $^2$Department of Economics, University of California, San Diego, California, United States\\
 \vspace{0.5cm}
}
		\maketitle

\onehalfspacing

\begin{abstract}
   Understanding treatment effect heterogeneity has become an increasingly popular task in various fields, as it helps design personalized advertisements in e-commerce or targeted treatment in biomedical studies. However, most of the existing work in this research area focused on either analyzing observational data based on strong causal assumptions or conducting post hoc analyses of randomized controlled trial data, and there has been limited effort dedicated to the design of randomized experiments specifically for uncovering treatment effect heterogeneity. In the manuscript, we develop a framework for designing and analyzing response adaptive experiments toward better learning treatment effect heterogeneity. Concretely, we provide response adaptive experimental design frameworks that sequentially revise the data collection mechanism according to the accrued evidence during the experiment. Such design strategies allow for the identification of subgroups with the largest treatment effects with enhanced statistical efficiency. The proposed frameworks not only unify adaptive enrichment designs and response-adaptive randomization designs but also complement A/B test designs in e-commerce and randomized trial designs in clinical settings. We demonstrate the merit of our design with theoretical justifications and in simulation studies with synthetic e-commerce and clinical trial data.
    
    \smallskip 
     
\end{abstract}

\onehalfspacing 

\section{Introduction}

\subsection{Motivation}

Understanding and characterizing treatment effect heterogeneity has become increasingly important in many scientific fields. For example, identifying differential treatment effects is an important step toward materializing the benefits of precision health, as it provides evidence regarding how groups of patients with specific characteristics respond to a given treatment either in efficacy or in adverse effects \cite{cross2029he}. As another example, individuals from different socioeconomic backgrounds may benefit differently from government programs, meaning that a careful evaluation of the program's possibly heterogeneous impacts is crucial for effective policy-making \citep{karlan2008credit,kharitonov2015sequential}. 

The existing literature in this research area mostly focuses on conducting retrospective analyses that employ observational or randomized experiment data. Even with large-scale observational data or carefully collected randomized experiment data, statistical bias in these analyses cannot be overlooked. On the one hand, in observational studies, statistical bias can arise due to potential violations of untestable causal assumptions \citep{athey2016recursive, athey2018approximate, cattaneo2019two, djebbari2008heterogeneous, hill2011bayesian, huang2012assessing, ma2020robust, raudenbush2015learning}. For example, one of the commonly imposed causal assumptions in practice is the unconfoundedness assumption, which states that conditional on measured confounders, the treatment assignment is as good as random \citep[see][for example]{glymour2016causal}. Due to the untestable nature of the unconfoundedness assumption and the possibility of unmeasured confounders in observational data, the validity of established causal conclusions under this assumption cannot be guaranteed \citep{imai2011unpacking, ma2020robust}. On the other hand, when analyzing classical randomized experiment data, although carrying out valid causal conclusion does not require imposing untestable causal assumptions, exploring treatment effect heterogeneity could still be susceptible to the winner's curse bias if seemingly promising heterogeneous treatment effects are selected from the data in an ad hoc fashion \citep{andrews2019inference,guo2021inference,ma2023breaking,stallard2008estimation}.  

In this manuscript, we tackle the problem of uncovering treatment effect heterogeneity from a fresh perspective by proposing an adaptive data collection mechanism called ``adaptive randomized experiments." This adaptive experiment approach enables the collection of reliable causal evidence that is specifically focused on understanding treatment effect heterogeneity. By adaptive, we mean that experimenters have the flexibility to sequentially allocate and modify experimental efforts, such as adjusting the treatment allocation probability and proportions of sequentially enrolled (sub)groups of participants. This adaptability allows experimenters to respond and make adjustments based on the evidence accrued during the experiment \cite{follmann1997adaptively, hu2006theory, rosenberger1993use, rosenblum2020optimal,russo2018tutorial,villar2015multi,villar2018response,xu2016subgroup}; see Section \ref{Sec:literature-review} for literature review and Section \ref{sec:overview} for a detailed introduction. 
To collect robust evidence towards learning treatment effect heterogeneity, we formalize our experimental design goal as maximizing the probability of correctly selecting subpopulations (or subgroups) who respond favorably to the treatment motivated by the large/moderate deviation principles (Section \ref{sec:objective}). Without loss of generality, we refer to the subpopulation with the highest treatment level as the best subpopulation or best subgroup throughout this manuscript.

\subsection{Our contribution}

In what follows, we break down our contributions from three perspectives:

Our proposed adaptive experiment strategy offers two potential benefits compared with the classical post hoc analysis approaches. First, because treatments are randomly assigned in adaptive experiments, they are independent of any potential unmeasured confounding variables, which means the proposed adaptive experiment design strategy generates samples enabling valid causal conclusions without imposing any untestable causal assumptions (see Section \ref{sec:overview} for our experimental setup). This is in stark contrast to analyses using observational data. Second, compared with conducting post hoc analyses with randomized experiment data, our design is equipped with the flexibility to revise the experimental strategy sequentially. Thus, the proposed adaptive experimental design strategy can detect individuals who respond favorably to the treatment and then optimize experimental effort spending based on the inferred context. As a part of this endeavor, compared to analyzing data collected from completely randomized experiments, this design feature offers advantages not only in improving the statistical efficiency of detecting treatment heterogeneity (Proposition \ref{proposition:comparison-cr})  but also in reducing the necessary sample size to correctly identify the best subgroup (Proposition \ref{proposition:sample-size-comparison}). 

Next, on the theoretical side, we first leverage the large and moderate deviation principle to characterize an ``oracle" allocation strategy, which maximizes the asymptotic probability of detecting the best subgroup when the underlying data generation process is unknown (Section \ref{sec:objective}). Because this oracle allocation strategy depends on the unknown data-generating process, the empirically feasible design strategy must be sequentially revised using adaptively collected data. 
As our second theoretical contribution, we demonstrate the oracle allocation is attainable using our proposed design strategies (Theorem \ref{theorem:allocation-consistency}). Unlike classical response adaptive randomization designs, we do not restrict the potential outcomes to following any parametric form, hence alleviating the burden of choosing what type of parametric assumptions should be used in practice. Our third contribution concerns the large-sample properties of the estimated treatment effects. Under mild moment restrictions, we show that the proposed design delivers asymptotically normally distributed estimators for subgroup treatment effects, and in particular, for the effect size of the best subgroup. In addition, we provide consistent asymptotic variance estimators and hence offer valid statistical inference procedures (Theorem \ref{theorem:normality}). 

We tackle three major challenges in our theoretical investigation, which arise as the data are sequentially collected, and hence, they are not statistically independent.  First, leveraging upon martingale methods, we develop a general framework and show that consistent estimation of features of the potential outcome distributions is possible for a large class of adaptive experimental designs (Lemma C.1 and its corollary in the Supplementary Materials), provided that the allocation probability and the enrichment proportion are bounded away from zero. Importantly, this result does not require independence and allows the data to be collected in an adaptive manner. More broadly, this technical lemma not only unifies adaptive enrichment designs and response adaptive randomization but also applies to other adaptive design settings with objective functions different from ours. We expect the general consistency result to be of independent interest. The second challenge in our theoretical investigation is that the optimized treatment assignment rule, both in finite samples and asymptotically, may not be unique, rendering standard M-estimation proof techniques inapplicable to our setting. Nevertheless, building on the general consistency result (Lemma C.1), we are able to show that with probability approaching one, any empirically optimized treatment assignment rule will be arbitrarily close to one of the oracle allocations (Lemma C.2 in the Supplementary Materials). Such ``set consistency'' result, seems new in the literature on adaptive experimental design. Finally, we establish the asymptotic normality of our treatment effect estimators. This result is established with a martingale central limit theorem \citep{hall2014martingale}, and hence, it takes into account the adaptive nature of the proposed experimental design and the induced dependence. In this process, a key step is to show that the cumulative empirical treatment probability (or the enrichment proportion) also converges to its oracle counterpart, which in turn helps to verify that a conditional variance will stabilize asymptotically. See Section C.7 in the  Supplementary Materials for details and additional discussions.

From a practical point of view, our proposed adaptive experiment strategy presents a unified framework incorporating both classical enrichment designs and response-adaptive randomization designs. Furthermore, our framework can be applied in both multi-stage and fully adaptive settings. See Table  \ref{table:design-strategy-simulation} for a summary of our design strategies. Thanks to its versatility, our designs can be applied to online experiments conducted in e-commerce platforms, clinical trials conducted in health industries, and policy evaluation experiments conducted for social science research.

\subsection{Existing literature}\label{Sec:literature-review}

Adaptive experiments have been frequently adopted in clinical trials where patients are enrolled sequentially based on certain eligibility criteria. In recent years, they also gained considerable popularity among online platforms for conducting A/B tests or digital randomized experiments. Our design strategy, proposed from a frequentist perspective, does not impose any parametric assumptions on the underlying data-generating process. Our setting is distinctly apart from Bayesian adaptive designs \cite{atkinson2005bayesian,cheng2005bayesian,gsponer2014practical, thall2007practical, park2022bayesian}.

Our framework falls into the realm of frequentist response adaptive designs. Such design strategies can be roughly divided into response-adaptive randomization (RAR) design and adaptive enrichment design in the existing literature. RAR design often refers to the design strategy in which the treatment assignment probabilities are adapted during the experiment based on the accrued evidence in the outcomes, with the goal of simultaneously achieving the experimental objectives and preserving statistical inference validity \citep{hu2006theory, rosenberger2002randomized, robertson2023response}.  The classical RAR framework often revises the treatment assignment probabilities at infinitely many stages, a design strategy we refer to as fully adaptive settings. In fully adaptive settings, a popular class of response-adaptive randomization designs is the doubly adaptive biased coin (DBCD) design. The early DBCD design can be found in \cite{eisele1994doubly}, which has its root in Efron's biased coin design \citep{efron1971forcing}. The asymptotic properties of DBCD designs are studied in various works \citep{hu2004asymptotic,hu2009efficient,tymofyeyev2007implementing, zhu2023seamless}. 
Our work shares some connections with \cite{tymofyeyev2007implementing} in that we both incorporate an optimization perspective into the problem of finding optimal treatment allocation, although our design objectives and the implementation of the optimal treatment allocation differ.
Other than the fully adaptive settings, existing RAR designs also accommodate multi-stage settings \citep{pocock1977group, van2008construction, zhao2023adaptive}. Such multi-stage designs propose to revise the treatment assignment probability by minimizing the asymptotic variance of the average treatment effect estimator. Nevertheless, this design is carried out in two stages and is not designed to identify treatment effect heterogeneity.

Instead of relying solely on the outcome variable to optimize for the experimental goals in RAR, one may further incorporate covariate information. RAR that further incorporate covariate information is known as covariate-adjusted response-adaptive (CARA) designs \citep{bandyopadhyay1999allocation,rosenberger2001covariate,zhang2007asymptotic, van2008construction}. Early work in \cite{zelen1994randomization} proposes to balance covariates based on the biased coin design. 
\cite{hu2015unified} propose a family of CARA designs that could account for both efficiency and ethics.  
 \cite{zhu2023covariate} generalizes CARA to incorporate semiparametric estimates. Some related CARA designs are also discussed in \cite{lin2015pursuit, villar2018covariate, zhao2022incorporating, zhao2023adaptive}. We note that much of the existing work on response-adaptive randomization designs and covariate-adjusted response-adaptive aims to optimize the estimation efficiency of the overall treatment effect but is not tailored to study treatment effect heterogeneity \citep{hu2003optimality,rosenberger2004maximizing}. 

On top of RAR, adaptive enrichment designs are often adopted in clinical trials, and interim data is used to identify treatment-sensitive patient subgroups by changing patient enrollment criteria. In these designs, experimenters often partition the population into pre-defined subgroups based on biomarkers measured at baseline and enroll patients in multiple stages \cite{wang2007approaches, rosenblum2011optimizing, rosenblum2014optimal, burnett2020adding}. For example, the early work in \cite{follmann1997adaptively} considers revising the enrollment proportions of two discrete patient subgroups defined by a single biomarker and provides conditions under which the type I error rate is controlled.
\cite{stallard2022adaptive} considers overlapping subgroups defined by a continuous biomarker. To our knowledge, different from our goal of identifying the best subgroup with high probability, much of the existing work on adaptive enrichment designs aims to preserve the type I error rate of the estimated subgroup treatment effect. 

As the data are sequentially collected using our design strategy, the toolkit we adopted to ensure the validity of statistical inference (martingale limit theories in particular) has also been used in the existing literature. For example, \cite{luedtke2016statistical, hadad2021confidence, zhan2021off, zhan2023policy} focus on analyzing adaptively collected data either from adaptive randomized experiments or online policy learning, different from our goal in designing an adaptive data collection mechanism.

Our work is connected to the literature on learning optimal policy. This area of literature is referred to by different names, such as choosing the best policy (e.g., \cite{kasy2021adaptive}) and identifying the best arm (e.g., \cite{russo2020simple}). Other related work includes \cite{audibert2010best}, \cite{garivier2016optimal}, and \cite{kaufmann2016complexity}. The differences between our work and theirs arise from the goals and constraints. In our work, the allocation ratios between control and treatment in different subgroups are determined by solving an optimization problem with the goal of maximizing the correct selection probability under the constraints of maintaining a non-zero propensity score and having limited resources to conduct the experiment. In contrast, \cite{kasy2021adaptive} adopt an allocation ratio between two arms that is asymptotically 1:1 directly enforced by an ``exploration sampling" constraint. In \cite{russo2020simple}, their top-two series algorithms by default have the allocation ratio between the ``top two" arms set as 1:1 and instead focus on using posterior probability to update which two arms are ``top two" on the fly. In our work, the optimal allocation ratio between two arms is not necessarily 1:1. Our work involves situations where implementing a treatment is costly, and there is an overall budget for how many treatments can be deployed. Such a budget constraint may impact the optimal allocation ratio. On the other hand, even with an unlimited budget, in our work, the optimal allocation ratio is not necessarily 1:1. Instead, it depends on the variances and relative mean gap between the two arms across different subgroups, which is rooted in our goal of maximizing the correct selection probability for the best subgroup. The differences in the goals and constraints between our work and the literature not only affect the treatment allocation and statistical methodology but also introduce additional methodological and technical challenges.

Furthermore, we would like to point out an additional delicate difference between our work and the broad learning and bandit literature. In most bandit problems, due to their prevalent applications within online recommendation, the most limited resource is a sample of users (experimentation unit), whereas the cost of various treatments is negligible. In our work, we consider situations where implementing the treatment can be costly per sample of a user (experimentation unit). For example, in clinical settings, randomized experiments can be expensive due to the high costs associated with treatment medication. 
Therefore, the resource constraints in our problem remain at the experimentation unit level and also at the level of how many treatments can be provided according to a budget.

Our work is also connected to the broad class of ``contextual bandit" problems (e.g., \cite{dudik2011efficient} and subsequent work). Contextual bandit problems aim to learn an optimal policy that maps a given covariate to the best arm for that covariate, where the metric is either to minimize cumulative regret or to minimize simple regret. In particular, the metric of minimizing a simple regret is connected to our metric of maximizing correct selection probability, although our goal is different from that of the contextual bandit problems. Instead of the goal of contextual bandit problems to identify the best arm for each subgroup (each covariate value), our work aims to identify the subgroup that benefits the most from a single new treatment under resource constraints. The optimization involved in solving for the optimal resource allocation is, therefore, different and requires different techniques to solve and analyze.

Lastly, while our work is connected to the classical design of experiments literature, we believe that the adaptive nature of our framework sets it apart. The early experimental designs can be traced back to \cite{fisher1936design}, which introduces the design principles such as blocking, randomization, and replication. The seminal work by \cite{wu2011experiments} lays down the foundation for diverse techniques and theories in experiment designs. For example, the orthogonal designs are a way to ensure that experiments yield clear, independent insights about each factor, thereby maximizing the information return on the experimental efforts and ensuring more reliable conclusions \cite{butler2001optimal,lin2010new,sun2017method}. As another example, with the advancement of computational capacity in modern designs, \cite{wu2001generalized} introduces a generalized minimum aberration criterion for evaluating asymmetrical factorial designs. A thorough review of factorial designs can be found in \cite{mukerjee2006modern} and the reference therein.

\section{A synthesized adaptive experiment framework}\label{sec:overview}

In this section,  we introduce a unified design framework in a two-arm (a treatment arm and a control arm) experiment along with notation. Our design framework operates within the frequentist framework. It encompasses classical response-adaptive randomization (RAR) design with adaptive enrichment (AE) design, both widely used in practice. 

Suppose experiment participants are sequentially enrolled in $T$ stages. The total number of enrolled subjects is $N = \sum_{t=1}^T n_t$, where $n_t$ denotes the number of subjects in Stage $t$, for $t=1,\ldots,T$. In Stage $t$, we denote the treatment assignment status of subject $i$ as $D_{it}\in\{0, 1\}$, where $i$ ranges from 1 to $n_t$. Here, $D_{it}=1$ corresponds to the treatment arm, while $D_{it}=0$ corresponds to the control arm. Denote subject $i$'s covariate information as $X_{it}\in\mathbb{R}^p$ and the observed outcome as $Y_{it}\in\mathbb{R}$. 

To formally introduce treatment (or causal) effects, we follow the Neyman-Rubin causal model \citep{neyman1923application, rubin1974estimating}. Define $Y_{it}(d)$ as the potential outcome we would have observed if subject $i$ receives treatment $d$ at Stage $t$, for $d\in\{0, 1\}$. 
The observed outcome can then be written as 
\begin{align*}
    Y_{it} = D_{it}Y_{it}(1)+(1-D_{it})Y_{it}(0), \quad i= 1, \ldots, n_t, \ t = 1, \ldots, T. 
\end{align*}
Consistent with the existing literature on adaptive experiments, we assume that the outcomes are observed without delay, and their underlying distributions do not shift over time \citep{hu2006theory}. Furthermore, we define the history, which represents the collected data up to Stage $t$, 
$$\bm{\mathcal{H}}_t = \{{\mathcal{H}}_s\}_{s=1}^t \triangleq \{(Y_{is}, D_{is}, X_{is}), i = 1, \ldots, n_s\}_{s=1}^t.$$

To investigate treatment effect heterogeneity, we partition the covariate sample space $\mathcal{X}$ into $m$ pre-specified non-overlapping regions, denoted as $\{\mathcal{S}_j\}_{j=1}^m$ (an extension of an overlapping division shall be discussed in Supplementary Materials Section G. In clinical settings, each partition of the sample space is commonly referred to as a subgroup \citep{assmann2000subgroup,kubota2014phase,xu2016subgroup}, where each subgroup comprises subjects with distinct characteristics. To evaluate the effectiveness of the treatment within each subgroup, we measure the mean difference between potential outcomes in the treated and control arms:
\begin{align*}
    \tau_{j} = \mathbb{E}[Y_{it}(1) - Y_{it}(0)|X_{it} \in \mathcal{S}_j],\quad t = 1,\ldots, T,\  j = 1,\ldots,m. 
 \end{align*}
Furthermore, we denote the total number of subjects enrolled in subgroup $j$ as $N_{j} = \sum_{t=1}^T n_{tj}$, where $n_{tj} = \sum_{i=1}^{n_t}\mathds{1}_{(X_{it}\in\mathcal{S}_j)}$.

In adaptive experiments, practitioners have the flexibility of sequentially allocating experimental efforts to reach certain pre-specified design goals. Such efforts include actively recruiting subjects of different characteristics in multiple stages and revising treatment assignment (or allocation) probabilities based on accrued evidence during the experiment. Within the existing literature, two commonly employed design strategies have emerged to distribute these experimental efforts differently, which we will discuss in detail below.

The first strategy is called response-adaptive randomization (RAR) design or covariate-adjusted response-adaptive (CARA) design. In these designs, experiments can sequentially revise the treatment assignment strategies based on responses accumulated during the experiment but, unlike enrichment designs, often do not change the enrollment criteria across multiple stages. RAR designs incorporating additional covariate information are more frequently referred to as covariate-adjusted response-adaptive (CARA) designs. The design goals of response-adaptive randomization designs tend to vary in different application areas, and we refer interested readers for \cite{robertson2023response} for a comprehensive review. Formally, by defining the treatment assignment probability (or propensity scores) for subjects in subgroup $j$ as 
\begin{align*}
    e_{tj} =\mathbb{P}(D_{it}=1|X_{it}\in\mathcal{S}_j), \quad t= 1,\ldots, T, \ j = 1,\ldots, m.
\end{align*}
RAR and CARA design aim to dynamically revise $e_{tj}$ to reach desired design goals. 

The second strategy is called (adaptive) enrichment design, which has been frequently carried out in clinical settings to identify patient subgroups that benefit the most from a given treatment \citep{follmann1997adaptively,lai2019adaptive,rosenblum2020optimal,simon2013adaptive}. In these designs, experimenters often fix the treatment allocation probability during the entire experiment, but they sequentially enroll different subgroups of participants over different stages. Here, the word ``enrichment" spells out the action of actively recruiting a new batch of subjects who may have characteristics different from the previous stage, and the word ``adaptive" indicates that the enrollment proportions of subjects with different characteristics can be adaptively revised based on the current understanding of treatment effect heterogeneity. Formally, by defining an auxiliary variable $Z_{it}\in\{1,0\}$ that indicates if subject $i$ is enrolled at Stage $t$, we introduce the enrichment proportion of subjects falling into region $\mathcal{S}_j$ in Stage $t$ as 
\begin{align*}
    p_{tj} = \mathbb{P}(X_{it}\in\mathcal{S}_j|Z_{it}=1), \quad t=1,\ldots, T, \ j=1,\ldots,m.
\end{align*}
Enrichment designs sequentially revise $p_{tj}$ across multiple stages to reach their design objectives.

Our proposed adaptive experimental design framework unifies response-adaptive randomization designs and enrichment designs by formalizing them as a sequential policy learning problem (see Table \ref{table:design-strategy-review} for a summary). We hope that this unified framework broadens the practicability of the proposed design framework under various practical constraints. In particular, we define a sequential policy $\bm{\pi}$ consisting of a sequence of policies $\pi_1, \ldots, \pi_{T-1}$, and  each $\pi_t$ is a mapping from the historical data $\bm{\mathcal{H}}_t = \{{\mathcal{H}}_s\}_{s=1}^t $ accumulated up to Stage $t$ to either the subgroup enrichment proportions $\bm{p}_{t+1} \triangleq (p_{t+1,1}, \ldots, p_{t+1,m})$, or to the treatment assignment probabilities $\bm{e}_{t+1} \triangleq (e_{t+1,1}, \ldots, e_{t+1,m})$, that is: 
\begin{align*}
  \pi_t:  \bm{\mathcal{H}}_t \rightarrow \bm{e}_{t+1} \triangleq (e_{t+1,1}, \ldots, e_{t+1,m}) & \quad\quad\quad\quad \text{Response-adaptive randomization design},\\
   \pi_t:  \bm{\mathcal{H}}_t \rightarrow \bm{p}_{t+1} \triangleq (p_{t+1,1}, \ldots, p_{t+1,m}) & \quad\quad\quad\quad \text{Adaptive enrichment design}.
\end{align*}
 
Other than dispending different experimental strategies, practitioners can also flexibly choose the number of stages $T$ and the number of participants $n_t$ in each stage of the experiment. We refer to experimental design strategies with large $n_t$ and finite $T$ as multi-stage designs, and we refer to designs with small $n_t$ and large $T$ as fully adaptive designs. While both designs tend to share similar large sample properties, they have different strengths and can often be applied in scenarios with different practical constraints. On the one hand, multi-stage designs can be preferable in clinical settings or social experiments where experimenters often have a limited number of opportunities to revise the experimental effort allocated during the experiment (see \cite{karlan2008credit,gertler2012investing} for example). Fully adaptive designs are more readily integrated into digital experiments such as online A/B testing or digital clinical trials in which sequentially allocating experimental efforts in a large number of stages is more practical and less costly (see \cite{kharitonov2015sequential,robertson2023response} for example). On the other hand, as seen in our simulation studies in Section \ref{sec:simulation}, benefiting from frequently updated experimental strategy, fully adaptive designs tend to have superior finite sample performance compared to multi-stage designs when the sample size $N$ is rather small. 

Benefiting from the above framework, while existing adaptive experiments normally target one of the experimental schemes listed in Table \ref{table:design-strategy-review},
the design strategies we shall propose can be applied in all four settings. This demonstrates that the proposed design strategy is flexible and completes existing frequentist adaptive design strategies, suggesting our designs can be potentially applied to online experiments conducted in e-commerce platforms, clinical trials conducted in health industries, and policy evaluation experiments conducted for social science research. In what follows, we introduce the general goal of our design strategy.

\begin{table}
\centering
\caption{Examples of frequentist data collection mechanisms in response adaptive experiments.\label{table:design-strategy-review}}
\begin{tabular}{c|c|c}
\hline\hline
&  {\textbf{(Regime 1)}}  & {\textbf{(Regime 2)}} \\
\multirow{2}*{$\pi_t$} & Small $n_t$ with large $T$  & Large $n_t$ with finite $T$  \\
&  {``Fully adaptive"}  & {``Multi-stage"} \\
\hline
& Response-adaptive randomization &  Adaptive propensity score   \  \\
\multirow{2}*{{Response-adaptive design}} & \cite{hu2006theory,rosenberger2002randomized, robertson2023response,eisele1994doubly,hu2004asymptotic,hu2009efficient,tymofyeyev2007implementing, zhu2023seamless} &  \cite{pocock1977group,zhao2023adaptive} \\
\multirow{2}*{ $\ \bm{\mathcal{H}}_t \rightarrow \bm{e}_{t+1}$} &  Covariate-adjusted & Sequential   \\
& response-adaptive  & rerandomization  \ \\ 
& \cite{bandyopadhyay1999allocation,rosenberger2001covariate,zhang2007asymptotic,hu2015unified,villar2018covariate,
zelen1994randomization,zhu2023covariate} & \cite{morgan2012rerandomization,morgan2015rerandomization,zhou2018sequential}  \\
\hline
{Enrichment design}  &    \multirow{2}*{Not available}   & Frequentist enrichment design \\
$\ \bm{\mathcal{H}}_t \rightarrow \bm{p}_{t+1}$ & &  \cite{wang2007approaches, rosenblum2011optimizing, rosenblum2014optimal, burnett2020adding,stallard2022adaptive,follmann1997adaptively}  \\
\hline\hline
\end{tabular}
\end{table}

\section{Design objectives and oracle allocation strategies: A large deviation perspective }\label{sec:objective}

Adaptive experiments are frequently designed with specific pre-determined goals in mind. Our adaptive experiment is designed with the goal of gathering strong evidence for learning treatment effect heterogeneity by identifying specific subgroups of participants who are more likely to benefit from the treatment.

Accurately identifying the best-performing subgroups provides several practical advantages, particularly in cases where one treatment is not universally beneficial for the entire population and treatment effects vary across different subpopulations. In clinical research, identifying the beneficial subgroup contributes to the development of personalized medicine, allowing treatments to be tailored to individual patients based on their unique characteristics or predictive markers. By identifying subgroups most likely to benefit, our trial design establishes the groundwork for targeted and individualized interventions. In social economics research, accurately identifying the best-performing subgroups enables policymakers and practitioners to understand which specific subpopulations are most positively affected by certain interventions or policies. This knowledge allows for more targeted and effective interventions to address social and economic challenges. By focusing resources and efforts on the subgroups that stand to benefit the most, policymakers can maximize the impact of their initiatives and improve overall societal well-being.

In statistical languages, our design goal is to construct reliable estimators of the subgroup average treatment effect so that the probability of correctly identifying the subgroups with the most beneficial (or harmful) effects is maximized when the experiment ends. Formally, without loss of generality, we assume that the population subgroup average treatment effects satisfy $\tau_1 > \tau_2 > \ldots > \tau_m$ (generalizations to other possible effect orders are provided in Supplementary Materials Section G), and suppose we have constructed consistent estimators $\hat{\tau}_1, \ldots, \hat{\tau}_m$ of $\tau_1, \ldots, \tau_m$ based on the collected data at the end of the experiment. Because the joint distribution of $\hat{\tau}_1, \ldots, \hat{\tau}_m$ not only depends on the underlying data distribution of the potential outcome and covariates but also crucially relies on the treatment assignment mechanism and subgroup enrollment proportions, these estimators can be viewed as a function of the historical data and the corresponding policy adopted in the adaptive experiment. Then, in a simple case where we aim to find the best subgroup with the largest treatment effect in the population (i.e., the first subgroup $\mathcal{S}_1$), our design objective is to find a sequential policy $\bm{\pi}$ belonging to a set of feasible policies $\bm{\Pi}$, so that the probability of the estimated first subgroup treatment effect margins out the others is maximized. As in this simple case, the first subgroup has the largest treatment effect in the population; the correction selection probability can be written as $ \mathbb{P}\big(\hat{\tau}_1 \geq \max_{2\leq j\leq m}\hat{\tau}_j\big)$. 

Unfortunately, without imposing additional parametric distributional assumptions on the historical data, directly searching for a policy that maximizes the correct selection probability results in an intractable optimization problem, as deriving a general analytic form of the correct selection probability is nearly impossible. One seemingly natural alternative is to consider solving this optimization problem in an asymptotic sense.  By letting the total sample size $N$ go to infinity, it is possible to approximate the distribution of $\hat{\tau}_j$ with a Gaussian distribution under mild conditions. However, even in this asymptotic framework, given $\tau_1 > \tau_2$ and for any policy $\pi$, the correct selection probability $\mathbb{P}(\hat{\tau}_1 \geq \max_{2\leq j\leq m}\hat{\tau}_j)$ grows exponentially fast to one as $N\rightarrow\infty$.  Consequently, it is no longer a function of $\bm{\pi}$, implying that directly searching for a sequential policy that maximizes the correct selection probability in an asymptotic sense is infeasible.

To address the challenges mentioned above, we temporarily shift our focus from studying a sequential policy that maximizes correct selection probability. Instead, we consider an idealized ``oracle" scenario in which we possess complete knowledge about the underlying data distribution. With this oracle in hand, we can explore the best strategy to allocate experimental efforts and design the experiment to achieve the highest possible correct selection probability. 

While acknowledging that this idealized scenario is not practically attainable, studying it can offer valuable insights and serve as a benchmark for evaluating the performance of more realistic strategies and policies in real-world adaptive experiments. However, even in the oracle scenario with perfect knowledge of the data distribution, the correct selection probability can still exhibit complex behavior with finite samples or tend to 1 as the sample size tends to infinity. Consequently, searching for the optimal allocation strategy remains a challenging task. In light of this, we are motivated to magnify the correction selection probability through the lens of the large and moderate deviation principle  \citep{dembo2009large, eichelsbacher2003moderate, hollander2000large, petrov2012sums}.

In essence, the large and moderate deviation principles provide a precise characterization of the correction selection probability using a set of rate functions. Specifically, under appropriate conditions with some $a_N\rightarrow\infty$ (as $N\rightarrow \infty$), the correct selection probability satisfies:

\begin{align}
  \nonumber   \lim_{N\rightarrow\infty} \frac{1}{a_N} \log \Big(1-\mathbb{P}\big( \hat{\tau}_1 \geq \max_{2\leq j \leq m} \hat{\tau}_j \big) \Big) & = - \min_{2\leq j\leq m}G(\mathcal{S}_1, \mathcal{S}_j; e_1, p_1, e_j, p_j) , \\
 G(\mathcal{S}_1, \mathcal{S}_j; e_1, p_1, e_j, p_j) & = \frac{(\tau_j - \tau_1)^2}{2\big( \mathbb{V}_1(e_1, p_1) +  \mathbb{V}_j(e_j, p_j)\big)} \label{eq:treatment-rate-function}, 
\end{align}
where $\mathbb{V}_j(e_j, p_j)$ is the variance of $\hat{\tau}_j$, for $j=1,\ldots, p$. The rate function $G(\mathcal{S}_1, \mathcal{S}_j;  e_1, p_1,  e_j, p_j)$ thus captures the exponential decay rate of the probability of the rare event where the estimated treatment effect in the best subgroup $\hat{\tau}_1$ is smaller than the estimated treatment effect in subgroup $\hat{\tau}_j$, as the sample size $N\rightarrow\infty$. The derivation of this result is explained in detail in Supplementary Materials Section H for mathematical clarity under both large and moderate deviation principles. Furthermore, depending on the design strategy, the rate function $G(\mathcal{S}_1, \mathcal{S}_j; e_1, p_1, e_j, p_j)$ typically has a closed-form expression that depends on the treatment allocations ($e_1$ and $e_j$) and subgroup enrichment proportions ($p_1$ and $p_j$) in the best subgroup $\mathcal{S}_1$ and subgroup $\mathcal{S}_j$; see \eqref{eq:rar-rate-variance} and \eqref{eq:enrichment-rate-function} for their closed-form expressions. 

Borrowing the language similar to \cite{donoho1994ideal} and \cite{fan2001variable}, we are ready to define \textit{oracle allocation strategies} in the
response-adaptive randomization designs and the enrichment designs as the following. In response-adaptive randomization designs, when the enrollment criteria are fixed, and the subgroup proportions cannot be modified, we define the oracle treatment allocation probabilities $\bm{e}^*\triangleq (e_1^*,\ldots,e_m^*)$ as the solution to the following constraint optimization problem:  
\begin{align*}
  \max_{\bm{e}} \Big\{\min_{2\leq j\leq m} G(\mathcal{S}_1, \mathcal{S}_j; e_1, e_j):\ \sum_{j=1}^m p_j e_j \leq c_1, \ c_2\leq e_j \leq 1-c_2\Big\}, 
\end{align*}
where $c_1\in (0,1)$ and $c_2\in (0,1/2)$. Similarly, in enrichment designs, when the treatment assignment probabilities in different subgroups are fixed, and the propensity scores $\bm{e} = (e_1, \ldots, e_m)$ cannot be modified, we define the oracle subgroup enrichment proportions $\bm{p}^*\triangleq (p_1^*,\ldots,p_m^*)$ as the solution to the following constraint optimization problem 
\begin{align*}
   \max_{\bm{p}} \Big\{\min_{2\leq j\leq m} G(\mathcal{S}_1, \mathcal{S}_j; p_1, p_j): \  \sum_{j=1}^m p_j = 1, \ p_j \geq 0\Big\}. 
\end{align*}
The closed-form solution of the above two optimization problems critically relies on the specific choice of the subgroup treatment effect estimators, and we thus leave more detailed discussions of the oracle allocation strategies for the response-adaptive randomization design and the enrichment design in Sections \ref{Sec:RAR} and \ref{sec:enrichment-design}. As shall be made clear in later sections, the oracle allocation strategies offer considerable advantages over traditional randomized experiments, including improving the efficiency in estimating the best subgroup treatment effect (Proposition \ref{proposition:comparison-cr}) and allowing the population treatment effect of the second-best subgroup to stay closer to that of the best subgroup (Proposition \ref{proposition:sample-size-comparison}). 

In practice, when experimenters have no prior knowledge about the joint distribution of the subgroup treatment effect estimators, adaptive experiments offer a natural environment to sequentially learn the unknown parameters in each subgroup and adjust the allocation of experimental efforts during the experiment. In the following sections, we aim to answer the following two research questions: When we have no prior information about the data-generating process, is it possible to carry out adaptive experimental design strategies that sequentially study the joint distribution of the underlying data and meanwhile use learned information to allocate experimental efforts better as the experiment progresses? When the experiment is finished, can such designs produce subgroup treatment effect estimators that have competing performances with the ones under the oracle allocation strategies?

\section{Response-adaptive randomization design with adaptive treatment allocation}\label{Sec:RAR}

In this section, we present the oracle treatment allocation strategy for response-adaptive randomization designs. Subsequently, we propose two design strategies for fully adaptive and multi-stage settings (refer to Table \ref{table:design-strategy-review}), both of which address the questions raised at the end of the previous section in a positive manner.

\subsection{Oracle treatment allocation in response-adaptive randomization designs}\label{Sec:oracle-proportion-RAR}

As the rate function depends on the choice of the subgroup treatment effect estimators, we adopt the inverse propensity score weighting (IPW) estimator with estimated propensity scores to estimate the subgroup treatment effects, that is 
\begin{align}\label{eq:ipw}
    \hat{\tau}_{j} =  \hat{\tau}_{Tj} = \frac{\sum_{t=1}^T\sum_{i=1}^{n_t} \mathds{1}_{(X_{it\in\mathcal{S}_j})} D_{it}Y_{it}}{\sum_{t=1}^T\sum_{i=1}^{n_t} \mathds{1}_{(X_{it\in\mathcal{S}_j})} D_{it}} - \frac{\sum_{t=1}^T\sum_{i=1}^{n_t} \mathds{1}_{(X_{it\in\mathcal{S}_j})} (1-D_{it})Y_{it}}{\sum_{t=1}^T\sum_{i=1}^{n_t} \mathds{1}_{(X_{it\in\mathcal{S}_j})} (1-D_{it})}, \quad j =1, \ldots, m.
\end{align}
We adopt this particular estimator as it is semiparametrically efficient, following results documented in \cite{hirano2003efficient}. We leave a discussion on the augmented IPW estimator \cite{robins1994estimation} to Supplementary Materials Section G. When the IPW estimator is adopted,  we are able to derive a closed-form expression of the rate function:
\begin{align}
    G(\mathcal{S}_1, \mathcal{S}_j; e_1, e_j) = \frac{(\tau_j - \tau_1)^2}{2\big( \mathbb{V}_1(e_1) +  \mathbb{V}_j(e_j)\big)},\quad 
    \mathbb{V}_j(e_j) = \frac{\sigma_j(1)^2}{p_je_j} + \frac{\sigma_j(0)^2}{p_j(1-e_j)},\label{eq:rar-rate-variance}
\end{align} 
where $ \mathbb{V}_j(e_j)$ is the  asymptotic variance of the estimator $\hat{\tau}_j$, and $\sigma_j(d)^2  = \Variance[Y(d)|X\in\mathcal{S}_j]$, $d=0,1$. We note that in response-adaptive randomization designs, the subgroup enrollment proportions $p_j$'s remain fixed throughout the experiment. Consequently, we denote the rate function and the asymptotic variance solely as functions of the subgroup treatment assignment probabilities $e_j$'s.

With the closed-form expression of the rate function in hand, we are now ready to explore the oracle treatment allocation $\bm{e}^*\triangleq (e_1^*,\ldots,e_m^*)$, which solves the following optimization problem:
\begin{small}
\leqnomode
    \begin{alignat}{3}\label{eq:optimization-problem-adaptive-treatment}
\tag*{\textbf{Problem $\bm{A}$}}
\hspace{1.7cm}\textbf{:} \
\max_{\bm{e}}&\ \min_{2\leq j\leq m}  \frac{(\tau_j - \tau_1)^2}{2\big( \mathbb{V}_1(e_1) + \mathbb{V}_j(e_j)\big)}, \ &&\leftarrow \text{Maximize correct selection probability}\\
        \text{s.t.}&\ \sum_{j=1}^m p_j e_j \leq c_1,  \ && \leftarrow \text{``Cost"/practical constraint} \nonumber \\ 
  &  \  c_2\leq e_j \leq 1-c_2,\ j =1, \ldots,m, \ && \leftarrow \text{Feasibility constraints} \nonumber
\end{alignat}
\end{small}
where $c_1\in (0,1)$ and $c_2\in (0, 1/2)$. Here, the cost/practical constraint restricts the proportion of subjects receiving the treatment, and the feasibility constraint restricts the treatment assignment probability in each subgroup to be bounded away from zero and one. 

Because the objective function is the minimum of $m-1$ rate function, the above optimization problem is nonlinear. We instead work with its equivalent epigraph representation:
\begin{small}
\leqnomode
    \begin{alignat}{3}\label{eq:adaptive-allocation-epigraph}
    \tag*{\textbf{Problem $\bm{B}$:}}
    \hspace{1.3cm} \
 \max_{\bm{e}}&\ z, \ && \leftarrow \text{Linear objective function for simple optimization}\\
        \text{s.t.}&\ \sum_{j=1}^m p_j e_j \leq c_1,  \ &&\leftarrow \text{``Cost"/practical constraint} \nonumber \\ 
  &  \  c_2\leq e_j \leq 1-c_2,\ j =1, \ldots,m, \ && \leftarrow \text{Feasibility constraints} \nonumber\\
  & \ \frac{(\tau_j - \tau_1)^2}{2\big(\mathbb{V}_1(e_1) + \mathbb{V}_j(e_j)\big)} - z\geq 0,\ j =2, \ldots,m. &&\leftarrow \text{Equivalent to maximize } \nonumber\\
   & \  &&\quad\text{ correct selection probability} \nonumber
\end{alignat}
\end{small}

\begin{figure}[h!]
    \centering
    \includegraphics[width=0.65\textwidth]{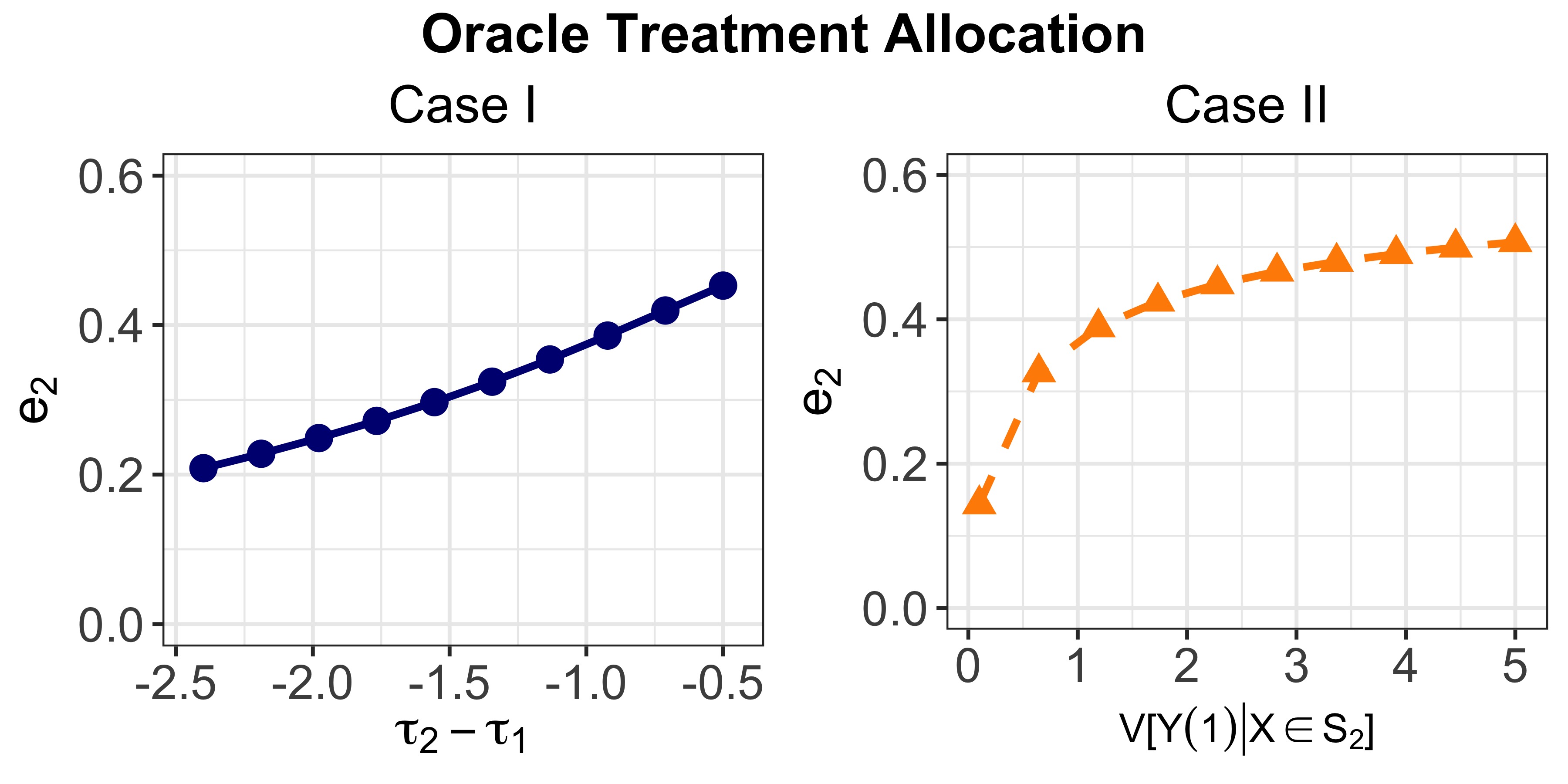}
    \caption{The change of oracle treatment allocation in the second subgroup in two different cases: (I) $\tau_1 = 3$, $\tau_3 = 0.5$, $\sigma_j(1)^2 = 2$, for $j=1,\ldots,3$, and (II) $\tau_1 = 3$, $\tau_2 = 2$, $\tau_3 = 1$, $\sigma_1(1)^2= \sigma_3(1)^2 = 2$.}
    \label{fig:oracle-allocation}
\end{figure}

The above epigraph representation has two key advantages. First, it formulates a concave optimization problem, enabling efficient solutions using open-source software such as \texttt{IPOPT} \citep{wachter2006implementation} and \texttt{GUROBI} \citep{gurobi2018gurobi}. Second, it facilitates exploration of the Lagrangian dual problem and allows us to obtain a simplified expression of the oracle treatment allocations in certain cases. For instance, suppose that the conditional variance of potential outcomes in the treatment and control arms is the same for each subgroup (that is, $\sigma_j(1)^2= \sigma_j(0)^2$), and assume that each subgroup has an equal enrollment proportion with $p_j = \frac{1}{m}$. In such cases, we can demonstrate that the oracle treatment allocation $\bm{e}^*= (e_1^*,\ldots,e_m^*)$ satisfies the following equation (see Supplementary Materials for the derivation):
\begin{align*}
    \frac{(\tau_j-\tau_1)^2}{\textcolor{black}{\frac{\sigma_1(1)^2}{e_1^*(1-e_1^*)}}+\frac{\sigma_j(1)^2}{e_j^*(1-e_j^*)}} = \frac{(\tau_k-\tau_1)^2}{\textcolor{black}{\frac{\sigma_1(1)^2}{e_1^*(1-e^*_1)}}+\frac{\sigma_k(1)^2}{e_k^*(1-e_k^*)}}, \quad j\neq k, \text{and } j, k \neq 1.
\end{align*}
This equation suggests that the required number of participants in the treatment arm of subgroup $j$ is reduced when it is relatively easier to distinguish subgroup $j$ from the best subgroup. This occurs when there is a larger difference between $\tau_j$ and $\tau_1$ or when the variance of subgroup $j$ is higher. To provide a clearer understanding, we consider a simple scenario with $m=3$. In Figure \ref{fig:oracle-allocation}, we plot the relationship between $e^*_2$, $\tau_2$, and $\sigma_2(1)^2$. 

Having obtained the oracle treatment allocation, we aim to approximate it using accrued data in an adaptive experiment. Next, we will discuss our proposed adaptive treatment allocation strategy in both fully adaptive and multi-stage settings.

\subsection{Fully adaptive case with large $T$ and small $n_t$}\label{subsec:adaptive-allocation-fully-adaptive}

In this section, we provide our proposed design strategy in the fully adaptive setting with large $T$ and small $n_t$ (Table \ref{table:design-strategy-review}). The derived fully adaptive response adjusted randomization  (RAR) sequential policy $\bm{\pi}_{\texttt{RAR}} = (\pi_1, \ldots, \pi_{T-1}) $ enables us to dynamically revise the treatment assignment probability so that the derived subgroup treatment effect estimator shares the same property as the one delivered by the oracle allocation strategies. 

\begin{stage1} Randomly assign subjects in each subgroup to the treatment arm with a pre-specified propensity score, such as $e_{1j} = \frac{1}{2}$, $j=1, \ldots, m$. 
\end{stage1}

As we have no prior information about enrolling participants, Stage $1$ of our design serves as an exploration stage. 
Note that in theoretical investigations, we allow Stage 1 sample size $n_1$ to be a vanishing fraction of the total sample size $N$, that is $\frac{n_1}{N}\rightarrow 0$. In practical implementations, we recommend enrolling at least $2$ subjects in each subgroup under each treatment arm. Therefore, $n_1 \geq 4\cdot m$. 

\begin{staget} Obtain $\hat{\bm{e}}^*_{t}$ by solving the sample analogue of \refeq{eq:adaptive-allocation-epigraph}, that is 
    \begin{align}\label{eq:adaptive-RAR}
 \hat{\bm{e}}^*_{t} = \arg \max_{\bm{e}}  \Big\{ z  :\ \sum_{l=1}^m \hat{p}_{tl} e_l \leq c_1, \ c_2\leq e_l \leq 1-c_2, \ \min_{2\leq j\leq m} \frac{( \hat{\tau}_{t-1,(j)} -  \hat{\tau}_{t-1,(1)})^2}{2\big(\hat{\mathbb{V}}_{t-1,(1)}(e_{1}) + \hat{\mathbb{V}}_{t-1,(j)}(e_{j})\big)} - z\geq 0 \Big\},
\end{align}
where the subscript $(j)$ indexes the subgroup with the $j$-th largest estimated treatment effect, and
\small{
\begin{align*}
    \hat{\tau}_{t-1,(j)} &=  
\frac{ \sum_{s=1}^{t-1} \sum_{i=1}^{n_s}\mathds{1}_{(X_{is}\in\mathcal{S}_{(j)})}D_{is}Y_{is}}{\sum_{s=1}^{t-1} \sum_{i=1}^{n_s}\mathds{1}_{(X_{is}\in\mathcal{S}_{(j)})}D_{is}}- \frac{ \sum_{s=1}^{t-1} \sum_{i=1}^{n_s}\mathds{1}_{(X_{is}\in\mathcal{S}_{(j)})}(1-D_{is})Y_{is}}{\sum_{s=1}^{t-1} \sum_{i=1}^{n_s}\mathds{1}_{(X_{is}\in\mathcal{S}_{(j)})}(1-D_{is})},\\ 
 \hat{\mathbb{V}}_{t-1,(j)}(e_j) &= 
\frac{\sum_{s=1}^{t-1}  
\sum_{i=1}^{n_s}\mathds{1}_{(X_{is}\in\mathcal{S}_{(j)})}D_{is}\Big(Y_{is} - \frac{ \sum_{s=1}^{t-1} \sum_{i=1}^{n_s}\mathds{1}_{(X_{is}\in\mathcal{S}_{(j)})}D_{is}Y_{is}}{\sum_{s=1}^{t-1} \sum_{i=1}^{n_s}\mathds{1}_{(X_{is}\in\mathcal{S}_{(j)})}D_{is}} \Big)^2}{\sum_{s=1}^{t-1}  \sum_{i=1}^{n_s} \mathds{1}_{(X_{is}\in\mathcal{S}_{(j)})}D_{is}} \Big({e_j \cdot \frac{\sum_{s=1}^{t-1}\sum_{i=1}^{n_s} \mathds{1}_{(X_{is}\in\mathcal{S}_{(j)})} }{\sum_{s=1}^{t-1}n_s  }}\Big)^{-1} \nonumber\\
&\quad+ \frac{\sum_{s=1}^{t-1}  \sum_{i=1}^{n_s}\mathds{1}_{(X_{is}\in\mathcal{S}_{(j)})}(1-D_{is})\Big(Y_{is} - \frac{ \sum_{s=1}^{t-1} \sum_{i=1}^{n_s}\mathds{1}_{(X_{is}\in\mathcal{S}_{(j)})}(1-D_{is})Y_{is}}{\sum_{s=1}^{t-1} \sum_{i=1}^{n_s}\mathds{1}_{(X_{is}\in\mathcal{S}_{(j)})}(1-D_{is})} \Big)^2}{\sum_{s=1}^{t-1}\sum_{i=1}^{n_s} \mathds{1}_{(X_{is}\in\mathcal{S}_{(j)})}(1-D_{is})}\\
&\quad\cdot \Big({(1-e_j) \cdot \frac{\sum_{s=1}^{t-1}\sum_{i=1}^{n_s} \mathds{1}_{(X_{is}\in\mathcal{S}_{(j)})} }{\sum_{s=1}^{t-1}n_s  }}\Big)^{-1} .   
\end{align*}}
Assign treatment with probability $\hat{\bm{e}}^*_{t}$.
\end{staget}

In each Stage $t$,  based on the newly collected data from the previous stage $\{\mathcal{H}_{t-1}\}$, we renew our understanding of the underlying data distribution and obtain a pair of updated estimates $(\hat{\tau}_{t-1,j},\hat{\mathbb{V}}_{t-1,j})$ for each subgroup.  These updated estimates thus enable us to better mimic the behavior of the oracle treatment allocation strategy by solving a refined optimization problem defined in Eq~\eqref{eq:adaptive-RAR} and revise the treatment assignment accordingly. 
We then assign treatment according to  $\hat{\bm{e}}^*_{t}$. 
In Stage $T$, we calculate empirical treatment assignment probabilities $\hat{\bm{e}}_T$ using the historical data collected up to Stage $T-1$. 

If Eq~\eqref{eq:adaptive-RAR} suggests multiple possible solutions, 
one can either choose a treatment allocation proportion that minimizes costs or a treatment allocation that is higher for the benefitted subgroup (i.e., a subgroup with a larger treatment effect). 

\begin{stageT}
Construct the final subgroup treatment effect estimator under the response-adaptive randomization (``RAR") design  along with its standard error using

\begin{align}\label{eq:ipw-final}
\hat{\tau}^{\texttt{RAR}}_{j} &= \frac{\sum_{s=1}^{T} \sum_{i=1}^{n_s}\mathds{1}_{(X_{is}\in\mathcal{S}_j)}D_{is}Y_{is}}{\sum_{s=1}^T\sum_{i=1}^{n_s} \mathds{1}_{(X_{is}\in\mathcal{S}_j)} D_{is} } -\frac{\sum_{s=1}^{T} \sum_{i=1}^{n_s}\mathds{1}_{(X_{is}\in\mathcal{S}_j)}(1-D_{is})Y_{is}}{\sum_{s=1}^T\sum_{i=1}^{n_s} \mathds{1}_{(X_{is}\in\mathcal{S}_j)} (1-D_{is}) }, \\ 
\hat{\mathbb{V}}^{\texttt{RAR}}_{j} &= 
\frac{\sum_{s=1}^{T}\sum_{i=1}^{n_s}\mathds{1}_{(X_{is}\in\mathcal{S}_j)}D_{is}\big(Y_{is} - \bar{Y}_{Tj}(1) \big)^2}{\sum_{s=1}^T \sum_{i=1}^{n_s} \mathds{1}_{(X_{is}\in\mathcal{S}_j)}D_{is}} \Big({\frac{\sum_{s=1}^{T}\sum_{i=1}^{n_s}  \mathds{1}_{(X_{is}\in\mathcal{S}_j)}D_{is}}{N}   } \Big)^{-1}  \label{eq:ipw-var-final}\\
&+
\frac{\sum_{s=1}^{T}\sum_{i=1}^{n_s}\mathds{1}_{(X_{is}\in\mathcal{S}_j)}(1-D_{is})\big(Y_{is} - \bar{Y}_{Tj}(0) \big)^2}{\sum_{s=1}^T \sum_{i=1}^{n_s} \mathds{1}_{(X_{is}\in\mathcal{S}_j)}(1-D_{is})} \Big({\frac{\sum_{s=1}^{T}\sum_{i=1}^{n_s}  \mathds{1}_{(X_{is}\in\mathcal{S}_j)}(1-D_{is})}{N}   } \Big)^{-1},\nonumber 
\end{align} 
where 
$\bar{Y}_{Tj}(d) = \frac{\sum_{s=1}^T \sum_{i=1}^{n_s}\mathds{1}_{(D_{is}=d)}\mathds{1}_{(X_{is}\in\mathcal{S}_j)}Y_{is}}{\sum_{s=1}^T \sum_{i=1}^{n_s} \mathds{1}_{(D_{is}=d)}\mathds{1}_{(X_{is}\in\mathcal{S}_j)}}$ for $d=0,1$.

Then, identify the best subgroup as the one exhibiting the maximal treatment effect size: 
\begin{align}\label{eq:j-star}
    j^*= \underset{1\leq j\leq m}{\argmax} \ \hat{\tau}^{\texttt{RAR}}_{j}. 
\end{align}
Lastly, construct a two-sided level-$\alpha$ confidence interval for the selected best subgroup as
\begin{align}\label{eq:CI}
    \Bigg[ \hat{\tau}^{\texttt{RAR}}_{j^*} \pm \Phi^{-1}(1-\alpha/2)\cdot \sqrt{\hat{\mathbb{V}}^{\texttt{RAR}}_{j^*} /N}\Bigg].
\end{align}
\end{stageT}

\subsection{Multi-stage case with small $T$ and large $n_t$}\label{subsec:treatment-allocation-multi-stage}

In this section, we provide an alternative multi-stage design strategy with small $T$ and large $n_t$, when experimenters can not revise the treatment assignment strategy too frequently. Stage 1 and the statistical inference after Stage $t$ are the same in fully adaptive and multi-stage settings. 
In Stage $t$, the multi-stage setting also requires an additional calibration step, as shown below: 

\begin{staget}
      (a) Solve for $\hat{\bm{e}}^*_{t}$ as in the fully adaptive setting. (b) In each subgroup, assign subjects to the treatment arm with probability $\tilde{e}_{t,(j)}$, where 
\begin{align*}
     \tilde{e}_{t,(j)}= 
\frac{\big(\hat{e}^*_{t,(j)}\sum_{s=1}^{t}\sum_{i=1}^{n_s} \mathds{1}_{(X_{is}\in\mathcal{S}_{(j)})}\big) - \sum_{s=1}^{t-1}\sum_{i=1}^{n_s} \mathds{1}_{(X_{is}\in\mathcal{S}_{(j)})}D_{is}}{\sum_{i=1}^{n_t}\mathds{1}_{(X_{it}\in\mathcal{S}_{(j)})}}, \ j = 1,\ldots, m.
\end{align*}
\end{staget}
By incorporating this additional step, we can ensure that the treatment allocation closely approximates the oracle treatment allocation. This adjustment is crucial because it enables the subgroup treatment effect estimators to compete with those obtained under the oracle allocation strategies. As a result, we can obtain accurate estimations of the treatment effects for different subgroups, even in scenarios where the oracle allocation is not directly feasible.

\section{Adaptive enrichment design}\label{sec:enrichment-design}

In this section, we start by introducing the oracle subgroup enrichment proportions. We then propose two enrichment design strategies for multi-stage and fully adaptive settings (Table \ref{table:design-strategy-review}).

\subsection{Oracle subgroup enrichment proportions}\label{subsec:oracle-enrichment}

In enrichment designs, the treatment allocation probabilities are often fixed; we thus denote the rate function as a function of subgroup enrollment proportions: 
\begin{align}\label{eq:enrichment-rate-function}
    G(\mathcal{S}_1, \mathcal{S}_j; p_1, p_j) &\nonumber= \frac{(\tau_j - \tau_1)^2}{2\big(\mathbb{V}_1(p_1) + \mathbb{V}_j(p_j)\big)}, \\
   \mathbb{V}_j(p_j) &= \frac{\sigma_j(1)^2}{p_je_j} + \frac{\sigma_j(0)^2}{p_j(1-e_j)},
\end{align}
where $\sigma_j(d)^2 = \Variance[Y(d)|X\in\mathcal{S}_j]$.
Note that although $\mathbb{V}_j(p_j)$ takes the same form as $\mathbb{V}_j(e_j)$ in Eq \eqref{eq:rar-rate-variance}, they carry rather different meanings. In adaptive enrichment designs, we keep $e_j$ fixed while adaptively changing enrichment proportion $p_j$; therefore, in Eq \eqref{eq:enrichment-rate-function}, the variance $\mathbb{V}_j$ is a function of $p_j$. 

The oracle subgroup enrichment proportion $\bm{p}^*\triangleq (p_1^*,\ldots,p_m^*)$ solves the following optimization problem (after transforming the original problem with the epigraph representation):
\begin{small}
\leqnomode
    \begin{alignat}{3}\label{eq:optimization-problem-adaptive-enrichment}
\tag*{\textbf{Problem $\bm{C}$}} 
\hspace{1.35cm}\textbf{:} \
\max_{\bm{e}}&\ z, \ && \leftarrow \text{Linear objective function}\\
        \text{s.t.}&\   \sum_{j=1}^m p_j = 1, \  p_j > 0,\ j = 1,\ldots, m, && \leftarrow \text{Feasibility constraints} \nonumber\\
  & \ \frac{(\tau_j - \tau_1)^2}{2\big(\mathbb{V}_1(p_1) + \mathbb{V}_j(p_j)\big)} - z\geq 0,\ j =2, \ldots,m. && \leftarrow \text{Equivalent to maximize} \nonumber\\
   & \  &&\quad\ \text{correct selection probability} \nonumber
\end{alignat}
\end{small}
The feasibility constraints require that the sum of subgroup enrichment proportions should be equal to 1, and each subgroup enrichment proportion must be non-negative. To provide some understanding of the oracle enrichment proportion $\bm{p}^*$, we once again present the closed-form expression of $\bm{p}^*$ in a simplified scenario (as discussed in Remark \ref{remark:enrichment-design}). However, it's important to note that our theoretical development does not rely on the assumptions made in this simplified setting.

\begin{remark}\label{remark:enrichment-design}(Oracle enrichment proportions in a simplified setting)
Supose $\mathbb{V}_1(p_1)$ is negligible compared to $\mathbb{V}_j(p_j)$, and let $\sigma^2_j = \sigma_j(1)^2/{e_j} + \sigma_j(0)^2/{(1-e_j)}$. 
The oracle enrichment proportion can be found through: 
   \begin{align}\label{eq:enrichment-proportion}
    p^*_{j} = \frac{\beta_{j}}{\sum_{j=1}^m \beta_{j}}, \ \beta_{j} = 
    \begin{cases}
   \sigma_{j}^2/(\tau_{j}-\tau_{1})^2 & \ j\neq 1,\\
     \sigma_{j}\sqrt{\sum_{l\neq 1} \beta_{l}^2/\sigma_{l}^2} & \ j=1.\\
    \end{cases}
\end{align} 
\end{remark}
By referring to Equation \eqref{eq:enrichment-proportion}, it becomes evident that the oracle enrichment proportion for subgroup $j$ is influenced by two factors: the variance within the subgroup and the squared distance between the treatment effect of subgroup $j$ and that of the best subgroup. In simpler terms, when the treatment effect $\tau_{j}$ of subgroup $j$ is closer to $\tau_{1}$ or when the subgroup's variance term $\sigma^2_{j}$ is higher, the enrichment proportion $p^*_{j}$ increases. Consequently, experimenters will enroll a larger proportion of subjects from subgroup $j$.

\subsection{Proposed adaptive enrichment designs}\label{subsec:adaptive-enrichment}

In this section, we illustrate our proposed enrichment design in the multi-stage setting with large $n_t$ and small $T$ (Table \ref{table:design-strategy-review}). The derived sequential policy enables us to adaptively revise subgroup enrichment proportions so that the estimated subgroup treatment effect estimator achieves a similar performance as the one obtained by the oracle enrichment strategy. 

\begin{stage1}
Enroll subjects from all the considered subgroups with equal proportions $p_{1j} = \frac{1}{m}$. Randomly the treatment with $e_{1j} = \frac{1}{2}$ for all enrolled subjects. 
\end{stage1}
We set equal enrichment proportions at the initial stage as this approach allows us to begin exploring each subgroup with an equal level of attention and resources.

\begin{staget}
(a) Obtain $\hat{\bm{p}}^*_{t}$ by solving for the sample analogue of \refeq{eq:optimization-problem-adaptive-enrichment}: 
\begin{align}\label{eq:optimization-enrichment}
     \hat{\bm{p}}^*_{t} = \arg \max_{\bm{p}}  \Big\{ z  :\ \sum_{l=1}^m p_{l} = 1, \ p_l > 0, \ l = 1,\ldots, m,\ \min_{2\leq j\leq m} \frac{( \hat{\tau}_{t-1,(j)} -  \hat{\tau}_{t-1,(1)})^2}{2\big(\hat{\mathbb{V}}_{t-1,(1)}(p_1) + \hat{\mathbb{V}}_{t-1,(j)}(p_j)\big)} - z\geq 0 \Big\},
\end{align}
where the subscript ${(j)}$ indexes the subgroup with the $j$-th largest  estimated treatment effect, and 
\small{
\begin{align*}
    \hat{\tau}_{t-1,(j)} &=  
\frac{ \sum_{s=1}^{t-1} \sum_{i=1}^{n_s}\mathds{1}_{(X_{is}\in\mathcal{S}_{(j)})}D_{is}Y_{is}}{\sum_{s=1}^{t-1} \sum_{i=1}^{n_s}\mathds{1}_{(X_{is}\in\mathcal{S}_{(j)})}D_{is}}- \frac{ \sum_{s=1}^{t-1} \sum_{i=1}^{n_s}\mathds{1}_{(X_{is}\in\mathcal{S}_{(j)})}(1-D_{is})Y_{is}}{\sum_{s=1}^{t-1} \sum_{i=1}^{n_s}\mathds{1}_{(X_{is}\in\mathcal{S}_{(j)})}(1-D_{is})},\\
  \hat{\mathbb{V}}_{t-1,(j)}(p_j) &=  \frac{\sum_{s=1}^{t-1} 
\sum_{i=1}^{n_s}\mathds{1}_{(X_{is}\in\mathcal{S}_{(j)})}D_{is}\big(Y_{is} -\frac{ \sum_{s=1}^{t-1} \sum_{i=1}^{n_s}\mathds{1}_{(X_{is}\in\mathcal{S}_{(j)})}D_{is}Y_{is}}{\sum_{s=1}^{t-1} \sum_{i=1}^{n_s}\mathds{1}_{(X_{is}\in\mathcal{S}_{(j)})}D_{is}}\big)^2}{\sum_{s=1}^{t-1} \sum_{i=1}^{n_s} \mathds{1}_{(X_{is}\in\mathcal{S}_{(j)})}D_{is} }\Big({\frac{\sum_{s=1}^{t-1}\sum_{i=1}^{n_s}\mathds{1}_{(X_{is}\in\mathcal{S}_{(j)})}D_{is}}{\sum_{s=1}^{t-1}\sum_{i=1}^{n_s}\mathds{1}_{(X_{is}\in\mathcal{S}_{(j)})}}\cdot p_j}\Big)^{-1} \\
+&  \frac{\sum_{s=1}^{t-1}
 \sum_{i=1}^{n_s}\mathds{1}_{(X_{is}\in\mathcal{S}_{(j)})}(1-D_{is}) \big(Y_{is}- \frac{ \sum_{s=1}^{t-1} \sum_{i=1}^{n_s}\mathds{1}_{(X_{is}\in\mathcal{S}_{(j)})}(1-D_{is})Y_{is}}{\sum_{s=1}^{t-1} \sum_{i=1}^{n_s}\mathds{1}_{(X_{is}\in\mathcal{S}_{(j)})}(1-D_{is})} \big)^2}{\sum_{s=1}^{t-1}\sum_{i=1}^{n_s} \mathds{1}_{(X_{is}\in\mathcal{S}_{(j)})} (1-D_{is})}\\
 &\cdot\Big({\frac{\sum_{s=1}^{t-1}\sum_{i=1}^{n_s}\mathds{1}_{(X_{is}\in\mathcal{S}_{(j)})}(1-D_{is})}{\sum_{s=1}^{t-1}\sum_{i=1}^{n_s}\mathds{1}_{(X_{is}\in\mathcal{S}_{(j)})}}\cdot p_j}\Big)^{-1}. 
\end{align*}}

(b) Enroll subjects from $\mathcal{S}_{(j)}$ with calibrated proportion $\tilde{p}_{t,(j)}$, where
\begin{align*}
    \tilde{p}_{t,(j)} = \frac{1}{n_t}\Big(\big(\hat{p}^*_{t,(j)}\sum_{s=1}^t n_s\big) - \sum_{s=1}^{t-1}\sum_{i=1}^{n_s}\mathds{1}_{(X_{is}\in\mathcal{S}_{(j)})}\Big), \ j = 1,\ldots, m.
\end{align*}
\end{staget}

Applying the same reasoning as discussed in Section \ref{subsec:treatment-allocation-multi-stage}, we calculate the estimated optimal subgroup enrichment proportion by subtracting the proportion of enrolled subjects in the previous stages. If Eq~\eqref{eq:optimization-enrichment} suggests multiple possible designs, 
one can choose an enrichment proportion that is closest to the uniform allocation of $1/m$ or a higher enrichment proportion for the subgroup with a larger treatment effect. 

\begin{stageT}
Construct the final subgroup treatment effect estimator $\hat{\tau}_j^{AE}$ and the variance estimator $\hat{\mathbb{V}}_j^{AE}$ under the adaptive enrichment (``AE")  design. These two estimators take the same form as in Eq \eqref{eq:ipw-final} and Eq \eqref{eq:ipw-var-final}.

Then, identify the best subgroup as the one exhibiting the maximal treatment effect size: 
$$j^* = \underset{1\leq j\leq m}{\argmax} \ \hat{\tau}^{\texttt{AE}}_{j}.$$
Lastly, construct a two-sided level-$\alpha$ confidence interval for the identified best subgroup $j^*$ as $$\Bigg[\hat{\tau}^{\texttt{AE}}_{j^*} \pm \Phi^{-1}(1-\alpha/2)\cdot \sqrt{\hat{\mathbb{V}}^{\texttt{AE}}_{j^*}/N}\Bigg].$$
\end{stageT}
We note that although $\hat{\tau}^{\texttt{AE}}_{j^*}$ shares the same 
form as $\hat{\tau}^{\texttt{RAR}}_{j^*}$, they follow different data generating distributions. This distinction shall be further illustrated in Theorem \ref{theorem:normality}.

\section{Theoretical investigation}\label{sec:theory}

In this section, we establish the theoretical properties of our proposed adaptive experiment design strategies. We start with introducing notations and assumptions in Section \ref{sec:notation-and-assumption}. We then provide a general result on the consistency of estimated moments of the potential outcomes in the adaptive setting (Lemma \ref{lemma:consistency of moments of potential outcomes} in Subsection \ref{subsection:some consistency results}), which encompasses the proposed designs as special cases. It is also worth mentioning that the consistency result applies to both fully adaptive ($T\to\infty$) and multi-stage scenarios ($T$ fixed), and hence it may be of independent interest. See the Supplementary Materials for additional results and discussions. Building on this lemma, we show in Subsection \ref{section: Theoretical results under our proposed adaptive experiment strategies} that (i) the proposed treatment allocation (in the RAR design) and enrichment proportion (in the AE design) converge to their oracle counterparts; (ii) both the actual treatment and enrichment frequencies converge asymptotically to the oracle values; (iii) in both RAR and AE settings, the estimated treatment effects are asymptotically normally distributed. Combined with a consistent variance estimator, results in Subsection \ref{section:Theoretical properties of the proposed adaptive experiment strategy} deliver a suite of estimation and statistical inference methods targeted at learning treatment effect heterogeneity with a rigorous statistical guarantee. In Section \ref{subsec:efficiency-gain}, 
we compare our proposed RAR design with the classical completely randomized design under simplified theoretical conditions. The theoretical comparison demonstrates three advantages of our design: (i) it corresponds to a smaller large deviation rate, suggesting a higher correct selection probability and a stronger estimation bias control (Theorem \ref{proposition:large-deviation-rate-comparison}); (ii) our design improves the statistical efficiency of uncovering treatment effect heterogeneity (Proposition \ref{proposition:comparison-cr}); (iii) our design reduces the required sample size for best subgroup identification (Proposition \ref{proposition:sample-size-comparison}). 

\subsection{Assumption and additional notation }\label{sec:notation-and-assumption}

We consider the asymptotic regime where the number of enrolled subjects, $N = \sum_{t=1}^T n_t$, grows, where we recall that $n_t$ is the number of subjects in Stage $t$. In the fully adaptive setting, this is equivalent to letting the number of time stages, denoted by $T$, approach infinity. In the multi-stage setting, $T$ is fixed, and $n_t$ will increase.  Additionally, we denote the total number of subjects enrolled in subgroup $j$ as $N_j= \sum_{t=1}^T n_{tj}$, which is the sum of subjects in subgroup $j$ across all stages.

We work under the following assumptions for our theoretical investigations: 

\begin{assumption}\label{assumption-1:boundedness}
(i) The potential outcomes and the covariates, $(Y_{it}(0), Y_{it}(1), X_{it})$, are independently and identically distributed across $t = 1, \ldots, T$ and $i=1,\ldots, n_t$. (ii) The potential outcomes have bounded fourth moments: $\mathbb{E}[|Y_{it}(d)|^{4}] < \infty$ for $d=0,1$. (iii) The potential outcomes have non-vanishing conditional variances: there exists some $\delta > 0$, such that $\mathbb{V}[Y_{it}(d)|X_{it}\in\mathcal{S}_{j}] \geq \delta$ for $d=0,1$ and $j=1,2,\dots,m$. 
\end{assumption}

\begin{assumption}\label{assumption-2:tau ranking}
There are $m \geq 2$ subgroups, and the subgroup treatment effects can be monotonically ordered with $\tau_1 >\tau_2 >\ldots > \tau_m$.
\end{assumption}

\noindent To simplify theoretical derivation, we assume that there are no exact ties among the population subgroup treatment effects. As an extension of our current framework, in the Supplementary Material Section G,  we provide a tentative approach to handle potential ties based on our earlier work \cite{wei2023inference}.

\begin{assumption}\label{assumption-3:subgroup-proportion positivity}
    (i) For the response-adaptive randomization (RAR) design, the subgroup proportions $p_1, \ldots, p_m$ are bounded away from 0 by a positive constant, that is, there exists a constant $\delta\in(0,1)$ such that $p_j \geq \delta $ for all $j$. (ii) For the adaptive enrichment (AE) design, the subgroup treatment assignment probabilities $e_1, \ldots, e_m$ are bounded away from 0 and 1; that is, there exists a constant $\delta \in (0, 1/2)$ such that $\delta \leq e_j \leq 1-\delta $ for all $j$.
\end{assumption}

\noindent Assumption \ref{assumption-3:subgroup-proportion positivity}, together with the constraints in our optimization problems A and C, ensures that each considered subgroup has a non-vanishing enrollment probability, and within each subgroup, participants will be assigned to both the treatment and control arms \cite{ma2020robust}. 

To facilitate theoretical discussions in the upcoming subsection, we will differentiate ``actual treatment allocations" and ``actual enrichment proportions" from those given by our algorithms. To be precise, the actual treatment allocations refer to the cumulative empirical treatment frequencies at each stage:
\begin{align*}
   \bm{\hat{e}}_t = \Big(\frac{\sum_{s=1}^t \sum_{i=1}^{n_s} \mathds{1}_{(X_{is}\in\mathcal{S}_1)}D_{is} }{\sum_{s=1}^t \sum_{i=1}^{n_s} \mathds{1}_{(X_{is}\in\mathcal{S}_1)}},\ \dots,\ \frac{\sum_{s=1}^t \sum_{i=1}^{n_s} \mathds{1}_{(X_{is}\in\mathcal{S}_m)}D_{is} }{\sum_{s=1}^t \sum_{i=1}^{n_s} \mathds{1}_{(X_{is}\in\mathcal{S}_m)}}\Big).
\end{align*}
We also adopt the convention $\bm{\hat{e}} = \bm{\hat{e}}_T$. Similarly, for adaptive enrichment designs, we define the actual enrichment proportions as:
\begin{align*}
   \bm{\hat{p}}_t = \Big( \frac{\sum_{s=1}^t \sum_{i=1}^{n_s} \mathds{1}_{(X_{is}\in\mathcal{S}_1)}}{\sum_{s=1}^t n_s},\ \dots,\ \frac{\sum_{s=1}^t \sum_{i=1}^{n_s} \mathds{1}_{(X_{is}\in\mathcal{S}_m)}}{\sum_{s=1}^t n_s}\Big),
\end{align*}
and $\bm{\hat{p}} = \bm{\hat{p}}_T$. In contrast, we use the terms ``optimized treatment allocations" and ``optimized subgroup enrichment proportions" to refer to the proposed design, solved from Eq~\eqref{eq:adaptive-RAR} and Eq~\eqref{eq:optimization-enrichment}:
\begin{align*}
  \bm{\hat{e}}^*_{t}= (\hat{e}^*_{t1},\ldots,\hat{e}^*_{tm}),\quad \text{and}\quad \bm{\hat{p}}^*_{t}= (\hat{p}^*_{t1},\ldots,\hat{p}^*_{tm}).
\end{align*}

\subsection{Theoretical properties of the proposed adaptive experiment strategy}\label{section:Theoretical properties of the proposed adaptive experiment strategy}

We are now ready to introduce the theoretical properties of our proposed adaptive experiment strategies. Subsection \ref{subsection:some consistency results} presents a general consistency result on the estimated moments of potential outcomes, which encompasses our proposed RAR and AE designs as special cases. 

\subsubsection{Consistency results in a general adaptive experiment setting}\label{subsection:some consistency results}

In the lemma below, we use the generic notation $\mathfrak{p}_{tj} = \bbp(X_{it}\in\mathcal{S}_j|\bm{\mathcal{H}}_{t-1})$ for the subgroup proportion, and $\mathfrak{e}_{tj} = \bbp(D_{it}=1|X_{it}\in\mathcal{S}_j,\bm{\mathcal{H}}_{t-1})$ for the treatment probability for subgroup $j$, where 
$\bm{\mathcal{H}}_{t-1}$ is the sigma-algebra formed by $(X_{is},D_{is},Y_{is})_{1\leq i\leq n_s,1\leq s\leq t-1}$ for $t=1,2,\dots,T$, and $\bm{\mathcal{H}}_0$ is the trivial sigma-algebra. The notations suggest that $\mathfrak{p}_{tj}$ and $\mathfrak{e}_{tj}$ can depend on $\bm{\mathcal{H}}_{t-1}$ hence allowing for adaptive designs. Notice that both the RAR and the AE designs are special cases: in the RAR design, we set $\mathfrak{p}_{tj} = p_j$ and $\mathfrak{e}_{tj} = \hat{e}^*_{tj}$; in the AE design, we have $\mathfrak{p}_{tj} = \hat{p}^*_{tj}$ and $\mathfrak{e}_{tj} = e_{j}$.

\begin{lemma}\label{lemma:consistency of moments of potential outcomes}
Assume Assumption \ref{assumption-1:boundedness} holds, and that there exists some $\delta \in (0,1/2)$ such that for all $j=1,2,\dots,m$ and $t= 1,2,\dots$, 
\begin{align*}
    \mathfrak{p}_{tj}\geq \delta,\text{ and } \delta \leq \mathfrak{e}_{tj} \leq 1-\delta.
\end{align*}
Then for any $j=1,2,\dots,m$, and any $t$ satisfying $\sum_{s=1}^{t}n_s\to\infty$,  
\begin{align*}
    \frac{\sum_{s=1}^t\sum_{i=1}^{n_s} 
\mathds{1}_{(X_{is}\in\mathcal{S}_j)}D_{is}Y_{is}^c}{\sum_{s=1}^t\sum_{i=1}^{n_s} 
\mathds{1}_{(X_{is}\in\mathcal{S}_j)}D_{is}} &= \bbe[Y_{it}(1)^c|X_{it}\in\mathcal{S}] + O_p\left(\frac{1}{\sqrt{\sum_{s=1}^t n_s}}\right),\\
    \frac{\sum_{s=1}^t\sum_{i=1}^{n_s} 
\mathds{1}_{(X_{is}\in\mathcal{S}_j)}(1-D_{is})Y_{is}^c}{\sum_{s=1}^t\sum_{i=1}^{n_s} 
\mathds{1}_{(X_{is}\in\mathcal{S}_j)}(1-D_{is})} &= \bbe[Y_{it}(0)^c|X_{it}\in\mathcal{S}] + O_p\left(\frac{1}{\sqrt{\sum_{s=1}^t n_s}}\right). 
\end{align*}
\end{lemma}

In addition to encompassing the RAR and AE designs as special cases, Lemma \ref{lemma:consistency of moments of potential outcomes} also applies to both fully adaptive and multi-stage settings. To be precise, fully adaptive corresponds to $T\to\infty$ and $n_t=1$ (or a fixed constant), in which case the consistency holds as $t\to \infty$. On the other hand, a multi-stage setting involves a fixed $T$. Then, the consistency result holds for each fixed $t$ as the cumulative sample size $\sum_{s=1}^{t}n_s$ tends to infinity.

A common challenge for proving consistency in adaptive experiments is that the treatment probability or the subgroup frequency can depend on historical data. Our proof strategy builds on explicit variance bounds, which is in contrast to the classical method that employs results on optional stopping   \cite{doob1936note, hu2004asymptotic} 

\subsubsection{Theoretical results under our proposed adaptive experiment strategies}\label{section: Theoretical results under our proposed adaptive experiment strategies}

In this subsection, we establish the theoretical properties of our proposed adaptive experimental design strategies, including (i) consistency of the optimized and actual treatment allocations for the RAR design, (ii) consistency of the optimized and actual enrichment proportions for the AE design; (iii) consistency and asymptotic normality of the estimated subgroup treatment effects. We also provide a consistent estimator for the asymptotic variance of the estimated treatment effects, which allows for valid statistical inference. To save space, we focus on the full adaptive setting ($T\to \infty$ and $n_t=1$). Similar results can be established for multi-stage designs (Regime 2 in Table \ref{table:design-strategy-review}), which we discuss in the Supplementary Materials.

To start, we show that the estimated variance is consistent as a function of the treatment probability in the RAR design or as a function of the subgroup proportion in the AE design.

\begin{corollary}\label{lemma:consistency of v-hat as a function of the pscore}
Assume Assumptions \ref{assumption-1:boundedness} and \ref{assumption-3:subgroup-proportion positivity} hold. Let $\delta\in(0,1/2)$ be some constant. Then
\begin{align*}
    \text{RAR design:}&\quad \sup_{\delta\leq e\leq 1-\delta}\Big|\hat{\bbv}_{tj}(e) - {\bbv}_j(e)\Big| = O_p\left(\frac{1}{\sqrt{\sum_{s=1}^t n_s}}\right),\\
    \text{AE design:}&\quad \sup_{\delta\leq p\leq 1-\delta}\Big|\hat{\bbv}_{tj}(p) - {\bbv}_j(p)\Big| = O_p\left(\frac{1}{\sqrt{\sum_{s=1}^t n_s}}\right),
\end{align*}
for $j=1,2,\dots,m$. 
\end{corollary}

Another useful corollary of Lemma \ref{lemma:consistency of moments of potential outcomes} is that the estimated subgroup treatment effects are consistent. 

\begin{corollary}\label{corollary:consistency of tau-hats}
Assume Assumptions \ref{assumption-1:boundedness} and \ref{assumption-3:subgroup-proportion positivity} hold. Then for both RAR and AE designs,
\begin{align*}
    \hat{\tau}_{tj} - \tau_j = O_p\left(\frac{1}{\sqrt{\sum_{s=1}^t n_s}}\right), \quad j = 1,\ldots, m. 
\end{align*}
If Assumption \ref{assumption-2:tau ranking} also holds, then as $t\to\infty$,
\begin{align*}
\mathbb{P}\big(\hat{\tau}_{t,(j)} =  \tau_j\big) \to 1,\qquad
\mathbb{P}\big({\tau}_{(j)} =  \tau_j\big) \to 1.
\end{align*} 
\end{corollary}

Building on Lemma \ref{lemma:consistency of moments of potential outcomes} and Corollary \ref{corollary:consistency of tau-hats}, we now present the theoretical properties of our proposed adaptive experiment strategies. As our proposed adaptive experiment strategies are derived by sequentially solving the optimization problems in Sections \ref{subsec:adaptive-allocation-fully-adaptive} and \ref{subsec:adaptive-enrichment}, we shall present Theorem \ref{theorem:allocation-consistency} which includes two related but conceptually different consistency results: the convergence of the optimized treatment allocation or enrichment proportion to their oracle values, and the consistency of the actual treatment allocation or enrichment proportion.

\begin{theorem}{\normalfont{(Asymptotic consistency of adaptive experiment strategies)}}\label{theorem:allocation-consistency}
Assume Assumptions \ref{assumption-1:boundedness}--\ref{assumption-3:subgroup-proportion positivity} hold. Assume that Problems A and C admit unique solutions.
Then for any $\delta > 0$ and as $t\to\infty$, for the optimized treatment allocation and enrichment proportion:

\begin{align*}
\text{RAR design:}&\quad \mathbb{P}\big(\big\Vert\hat{\bm{e}}_{t}^*-\bm{e}^*\big\Vert\leq\delta\big)\rightarrow 1, \\
\text{AE design:}&\quad \mathbb{P}\big(\big\Vert\hat{\bm{p}}_{t}^*-\bm{p}^*\big\Vert\leq\delta\big)\rightarrow 1.
\end{align*}
In addition, for the actual treatment allocation and enrichment proportion:
    \begin{align*}   
    \text{RAR design:}&\quad\mathbb{P}\big(\big\Vert\hat{\bm{e}}_t-\bm{e}^*\big\Vert\leq\delta\big)\rightarrow 1,\\
    \text{AE design:}&\quad\mathbb{P}\big(\big\Vert\hat{\bm{p}}_t-\bm{p}^*\big\Vert\leq\delta\big)\rightarrow 1.
    \end{align*}
    
\end{theorem}

The first part of Theorem \ref{theorem:allocation-consistency} suggests that the empirically and sequentially optimized treatment allocations and enrichment proportions converge to their oracle counterparts. 
We assume that the optimization problems admit unique solutions in Theorem \ref{theorem:allocation-consistency} because in practical implementations, non-uniqueness of the solution is not a serious concern in our setting. Whenever Eq \eqref{eq:adaptive-RAR} produces multiple treatment allocations, the researcher can always choose one using some additional criteria (say, with the smallest cost). We provide more details of selecting the set of optimizers in Sections \ref{subsec:adaptive-allocation-fully-adaptive} and \ref{subsec:adaptive-enrichment}.
For this reason, we are able to assume
that the solution to the optimization problems is unique throughout the rest of the paper.

The second part of Theorem \ref{theorem:allocation-consistency} implies that the actual treatment allocations ---  the fraction of subjects assigned to receive treatment in each subgroup --- converge to the oracle treatment allocation rule. The same consistency result holds for the enrichment design, as the actual subgroup proportions will converge to their oracle counterparts. In other words, although our proposed designs in Sections \ref{subsec:adaptive-allocation-fully-adaptive} and \ref{subsec:adaptive-enrichment} have no prior knowledge about the underlying data distribution before the experiment starts, they can allocate experimental efforts in a similar fashion to the oracle strategies when the sample size is sufficiently large.

\begin{theorem}{\normalfont{(Asymptotic normality and consistent variance estimation)}}\label{theorem:normality}
Assume Assumptions \ref{assumption-1:boundedness}--\ref{assumption-3:subgroup-proportion positivity} hold. In addition, assume that Problems A and C admit unique solutions, which are denoted by $\bm{e}^*$ and $\bm{p}^*$. Then as $t\to \infty$,
\begin{align*}
\text{RAR design:}&\quad \sqrt{N}\big(\hat{\tau}^{\texttt{RAR}}_{j^*}- \tau_1\big) \overset{\mathcal{D}}{\to} \mathcal{N}\big(0, \ \mathbb{V}_1(e^*_1)\big),\quad \mathbb{V}_1(e^*_1) = \frac{\sigma_1(1)^2}{p_1e^*_1} + \frac{\sigma_1(0)^2}{p_1(1-e^*_1)},\\
\text{AE design:}&\quad \sqrt{N}\big(\hat{\tau}^{\texttt{{AE}}}_{j^*}- \tau_1\big) \overset{\mathcal{D}}{\to} \mathcal{N}\big(0, \ \mathbb{V}_1(p^*_1)\big),\quad \mathbb{V}_1(p^*_1) = \frac{\sigma_1(1)^2}{p^*_1e_1} + \frac{\sigma_1(0)^2}{p^*_1(1-e_1)}.
\end{align*}
In addition,
\begin{align*}
    {\hat{\mathbb{V}}}_{j^*}^{\texttt{RAR}}  - \mathbb{V}_1(e_1^*) &= O_p\Big({\frac{1}{\sqrt{N}}}\Big),\quad \hat{\mathbb{V}}^{\texttt{AE}}_{j^*}  - \mathbb{V}_1(p_1^*) = O_p\Big(\frac{1}{\sqrt{N}}\Big).
\end{align*}
\end{theorem}
The theoretical results established in Theorem \ref{theorem:normality} indicate that the selected best subgroup treatment effect is a $\sqrt{N}$-consistent estimate of the best subgroup treatment effect $\tau_1$. In addition, the asymptotic variance can be consistently estimated by $\hat{\mathbb{V}}_{j^*}$.  This further suggests that the constructed confidence interval for the best subgroup, as given by Eq \eqref{eq:CI}, has correct coverage asymptotically. The asymptotic normality result relies on the martingale central limit theorem \citep{hall2014martingale} and the consistency results of our proposed adaptive experiment strategies. For its formal proof, we refer readers to the Supplementary Materials.

\subsection{Comparison with completely randomized experiments}\label{subsec:efficiency-gain}

In this section, we compare our proposed RAR design with completely randomized experiments, where the treatment is randomly assigned with a pre-fixed probability throughout the entire experiment. To simplify theoretical derivations, we work under the assumption that the outcome variables follow Gaussian distributions and the treatment assignments are independent, enabling us to conveniently compare the large deviation rates between our design and complete randomization. Concretely, the comparisons will be examined from three perspectives: (1) the large deviation rate and estimation bias (Proposition \ref{proposition:large-deviation-rate-comparison}), (2) the asymptotic variance of the estimated best subgroup treatment effect (Proposition \ref{proposition:comparison-cr}), and (3) the minimum sample size required to achieve a predetermined correct selection probability (Proposition \ref{proposition:sample-size-comparison}).

In order to establish a fair comparison with completely randomized experiments, we employ the same IPW estimator with estimated propensity scores to estimate the treatment effect, denoted as:
\begin{align*}
    \hat{\tau}_j^{\texttt{CR}} &=  
   \frac{\sum_{i=1}^{N}\mathds{1}_{(X_{i}\in\mathcal{S}_j)}D_{i}Y_{i}}{\sum_{i=1}^{N}  \mathds{1}_{(X_{i}\in\mathcal{S}_j)}D_i} - \frac{\sum_{i=1}^{N}\mathds{1}_{(X_{i}\in\mathcal{S}_j)}(1-D_{i})Y_{i}}{\sum_{i=1}^{N} \mathds{1}_{(X_{i}\in\mathcal{S}_j)}(1-D_i)},
\end{align*}
for $j=1,\ldots,m$. In this section, we consider a setting where (1) subgroup proportions are equal: $p_1=p_2=\ldots=p_m = \frac{1}{m}$, and (2) there exists a cost constraint: $\sum_{j=1}^m p_j e_j\leq c_1$, $c_1\in (0,1)$. 
In the completely randomized design, we set $\hat{e}_{tj}^* = c_1$ for every $t$ and $j$. This ensures that this design is comparable to ours while also meeting the cost constraint. The variance of $  \hat{\tau}_j^{\texttt{CR}}$ can be derived with a simple form:
\begin{align}\label{eq:SPEB}
\mathbb{V}_{j}(c_1) = \frac{\sigma_j(1)^2}{p_j\cdot c_1} + \frac{\sigma_j(0)^2}{p_j\cdot c_1}. 
\end{align}

We start by comparing the large deviation rates under the proposed response-adaptive randomization design and the complete randomization design. 

\begin{proposition}[Large deviation rate comparison]\label{proposition:large-deviation-rate-comparison}
Under Assumptions \ref{assumption-1:boundedness}-\ref{assumption-3:subgroup-proportion positivity}, 
\begin{align*}
 \lim_{N\rightarrow \infty} \frac{1}{N} \log \  (1-\mathbb{P}(\hat{\tau}^{\texttt{RAR}}_1\geq \max_{2\leq j\leq m}\hat{\tau}^{\texttt{RAR}}_j)) \leq 
 \lim_{N\rightarrow \infty} \frac{1}{N} \log \  (1-\mathbb{P}(\hat{\tau}^{\texttt{CR}}_1\geq \max_{2\leq j\leq m}\hat{\tau}^{\texttt{CR}}_j)),
\end{align*}
where $\hat{\tau}^{\texttt{CR}}_{1}$ denotes the estimated treatment effect of the best subgroup under the complete randomization design. 
\end{proposition}

\noindent Note that under Assumption \ref{assumption-2:tau ranking}, the correct selection probability $\mathbb{P}(\hat{\tau}^{\texttt{RAR}}_1\geq \max_{2\leq j\leq m}\hat{\tau}^{\texttt{RAR}}_j)$ can be equivalently written as $\mathbb{P}(j^* = 1)\rightarrow 1$, where $j^*= \underset{1\leq j\leq m}{\argmax} \ \hat{\tau}^{\texttt{RAR}}_{j}$ is the index of the selected best subgroup, as defined in  \eqref{eq:j-star}. 

\begin{figure}[h!]
    \centering
\includegraphics[width=0.85\textwidth]{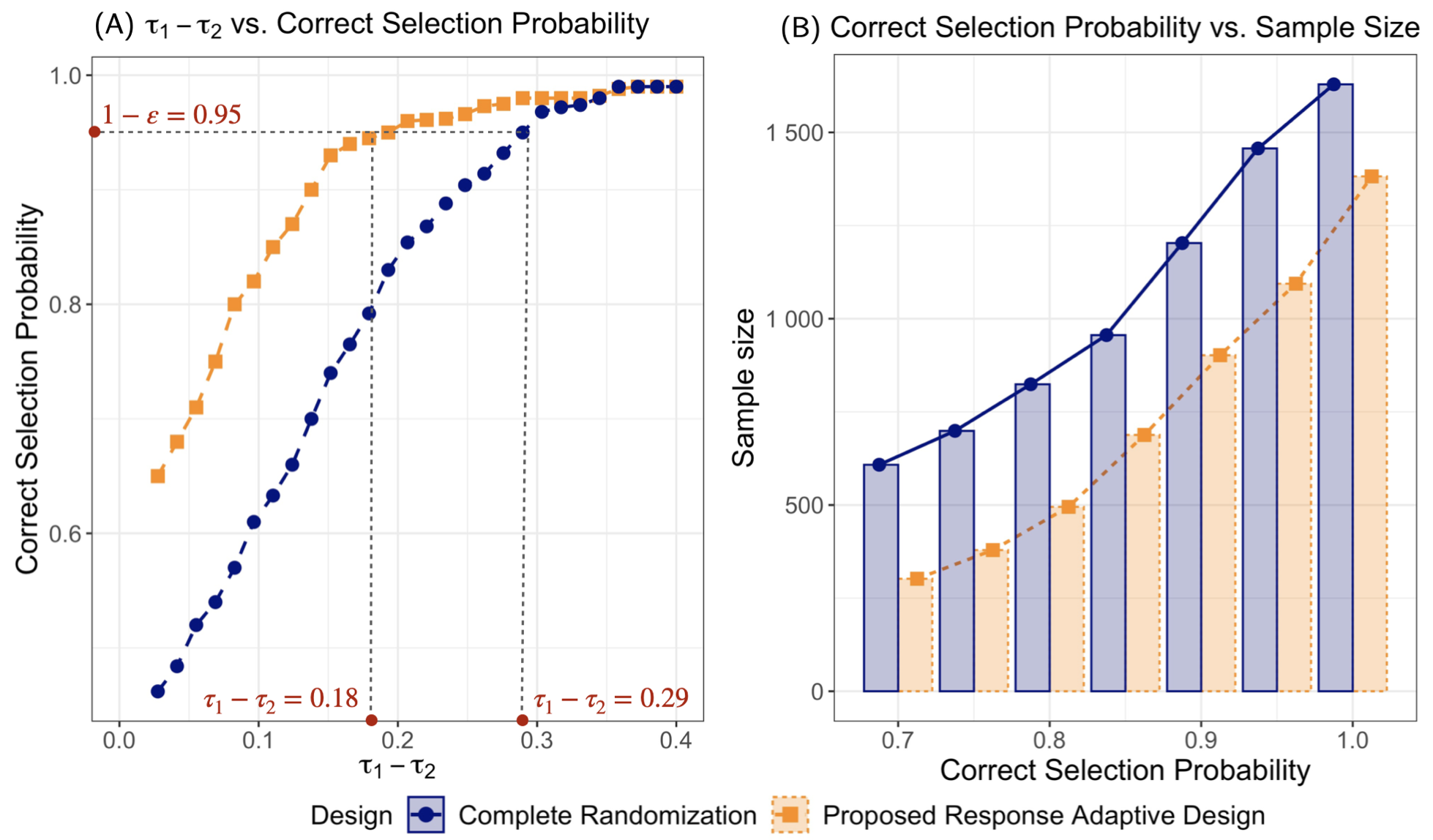}
    \caption{Verification of Propositions \ref{proposition:large-deviation-rate-comparison} and \ref{proposition:sample-size-comparison}. (A) Correct selection probability comparison
    with respect to various distances between $\tau_1$ and $\tau_2$. (B) Sample size comparison between the proposed oracle response-adaptive randomization design and the complete randomization design when fixing $\tau_1-\tau_2=0.1$.  }
    \label{fig:tau1tau2-comparison}
\end{figure}

Proposition \ref{proposition:large-deviation-rate-comparison} suggests that our proposed RAR design has a faster large deviation rate than that obtained under completely randomized experiments (i.e., the rate function implied by our method is larger in magnitude). This result has two indications. First, it implies that the probability of correctly selecting the best subgroup in our RAR design converges to one exponentially faster than in complete randomization as the sample size increases; see Figure \ref{fig:tau1tau2-comparison}(A) for verification of Proposition \ref{proposition:large-deviation-rate-comparison}. There, we provide a simulation study with a fixed sample size $N=500$ and set $\tau_1 = 1.6$, $\tau_3 = 0.5$, and $\tau_2 = \tau_1 - \delta$, where $\delta \in \{0.03,0.04,\ldots,0.4\}$. We compare the correct selection probability under the proposed design and the complete randomization design with respect to various distances between $\tau_1$ and $\tau_2$. 
Furthermore, a faster large deviation rate indicates that our design provides stronger bias control of the selected best subgroup compared to complete randomization. This is because the estimation bias of the best subgroup is proportional to the incorrect selection probability of the best subgroup, as shown in the following equation: 
\begin{align*}
\mathbb{E}[\hat{\tau}_{j^*}] - \tau_1 = -\tau_1\cdot \underbrace{\big(1-\mathbb{P}(\hat{\tau}_{1}\geq \max_{2\leq j\leq m} \hat{\tau}_j)\big)}_{\text{incorrect selection prob.}}. 
\end{align*}
Our design achieves stronger control over the incorrect selection probability, which in turn allows for better bias regulation compared to complete randomization. This conclusion can also be verified through our simulation results in Figure \ref{fig:adaptive-treatment-fs}(B).

Next, we compare the asymptotic efficiency gain of the proposed design for estimating the best subgroup treatment effect with the complete randomization design. Note that both variance lower bounds derived from our proposed design in Eq \eqref{eq:rar-rate-variance} and the complete randomization design in Eq \eqref{eq:SPEB} share a similar form, which allows us to compare the performance of our design with complete randomization.
To provide some insights into the efficiency comparison, we consider a simple case formalized in Proposition \ref{proposition:comparison-cr} below. In Supplementary Materials Section D.6, we consider more general settings and provide additional theoretical insights therein.

\begin{proposition}[Asymptotic variance comparison]\label{proposition:comparison-cr}
Assume (1) $\sigma_j(1)^2 = \sigma_j(0)^2$,  for $j=1,\ldots,m$, (2) $\sigma_1(1)^2 = \ldots = \sigma_m(1)^2 $, and (3) the cost constraint $c_1 < 0.5$. For all possible oracle treatment allocations $\bm{e}^* = (e_1^*, \ldots, e_m^*)\in \mathcal{E}^*$, we have for $j=1, \ldots, m$, 
\begin{align*}
\begin{cases}
   \mathbb{V}_{j}(e_j^*)\leq \mathbb{V}_{j}(c_1) , \quad 
   &\text{if}\  (\tau_j-\tau_1)^2 - \frac{1}{e_1^*(1-e_1^*)} \leq \frac{1}{c_1^2},\\
   \mathbb{V}_{j}(e_j^*) > \mathbb{V}_{j}(c_1) , \quad &\text{if}\  (\tau_j-\tau_1)^2- \frac{1}{e_1^*(1-e_1^*)}>\frac{1}{c_1^2}.
\end{cases}
\end{align*}
\end{proposition}

Proposition \ref{proposition:comparison-cr}  shows the efficiency comparison between our proposed RAR design and complete randomization. When estimating the best subgroup treatment effect, the asymptotic variance under our proposed RAR design is smaller than the complete randomization design. However, when $\tau_j$ is far away from $\tau_1$ or when the expected variance of the outcome in subgroup $j$ is small, our proposed response-adaptive randomization design is less likely to have efficiency gain. Proposition \ref{proposition:comparison-cr} thus entails the efficiency trade-off between our proposed design and the complete randomization design.  The efficiency trade-off can also be seen in Figure \ref{fig:real-data}.  In Supplementary Materials Section D.6, we provide another result without restricting $c_1 < 0.5$ and all variance terms to be equal.

Lastly, we compare the minimum sample size required to reach a fixed correct selection probability level.

\begin{proposition}[Sample size comparison]\label{proposition:sample-size-comparison}
Assume (1) $\sigma_j(1)^2 = \sigma_j(0)^2 $ for $j=1,\ldots,m$, (2) $\sigma_1(1)^2 = \ldots = \sigma_m(1)^2 $, and (3) the cost constraint $c_1 < 0.5$.  Suppose we aim to reach a correct selection probability of at least $1-\varepsilon$. For some positive constants $C < \infty$ and $C' < \infty$, 
    under the complete randomization design, the required sample size is characterized as 
    \begin{align*}
        N &\geq \frac{1}{(\tau_1-\tau_2)^2}\cdot\Big|C\cdot \log(\varepsilon) \cdot \sigma_1(1)^2\Big|.
    \end{align*}
Under our proposed response-adaptive randomization design, for all possible oracle treatment allocations $\bm{e}^* = (e_1^*, \ldots, e_m^*)\in \mathcal{E}^*$, the required sample size is characterized as 
     \begin{align*}
       N  &\geq \Big(\frac{1}{(\tau_1-\tau_2)^{2}}\cdot \Big|C' \cdot \log(\varepsilon)\cdot \sigma_1(1)^2\Big|\Big)^{3/4}.
    \end{align*}
\end{proposition}

Proposition \ref{proposition:sample-size-comparison} says that to reach a correct selection probability level of at least $1-\varepsilon$, our proposed adaptive design strategy often requires a smaller sample size compared to the complete randomization design. To verify Proposition \ref{proposition:sample-size-comparison}, we provide a simple simulation in Figure \ref{fig:tau1tau2-comparison}(B). 
In Figure \ref{fig:tau1tau2-comparison}(B), we fix $\tau_1-\tau_2=0.1$ and investigate the sample size needed to reach various correct selection probability levels. Figure \ref{fig:tau1tau2-comparison}(B) demonstrates that to reach a prespecified correct selection probability level, our proposed design requires smaller sample sizes.  In other words, when $\tau_1$ is close to $\tau_2$, our proposed response-adaptive randomization design correctly distinguishes the best subgroup from the second best subgroup with a higher probability.

\section{A synthetic case study}\label{sec:simulation}

In this section, we investigate the performance of our proposed response-adaptive randomization design and adaptive enrichment design in a synthetic case study using e-commerce data. We summarize four takeaways as follows: First, compared to several classical experimental design strategies, our proposed design requires the smallest sample size to reach a pre-fixed level of correct selection probability (Panel (A) in Figures \ref{fig:adaptive-treatment-fs}--\ref{fig:adaptive-enrichment-ms}). Benefiting from an improved correct selection probability, our design also yields the lowest estimation bias for the best subgroup (Panel (B) in Figures \ref{fig:adaptive-treatment-fs}--\ref{fig:adaptive-enrichment-ms}). Second, our proposed response-adaptive randomization design yields a smaller variance when estimating the best subgroup treatment effect (Figures \ref{fig:real-data}--\ref{fig:adaptive-treatment}). Third, the fully adaptive setting achieves an equivalent correct selection probability with less experimental data compared to the multi-stage setting, while the multi-stage setting can be more practical to implement as it requires fewer updates (Figure \ref{fig:adaptive-treatment-fs}A versus Figure \ref{fig:adaptive-treatment}A).

\subsection{Synthetic case study background}

We design our synthetic case study using e-commerce data collected from ModCloth, a website specializing in women's apparel. A crucial marketing strategy for apparel-based websites is the use of human models to showcase their products. Various studies indicate a prevailing ``pro-thin" bias in fashion advertising, suggesting that such websites often tend to display idealized, size-small models wearing their clothes \citep{levine2015affective,aagerup2011influence}. However, in light of the recent social campaigns advocating for inclusiveness in fashion marketing, some fashion companies have revised their advertising strategies to feature models of a wider range of body shapes \citep{cinelli2016role}. While it is hypothesized that the inclusive advertising strategy could improve customer satisfaction, it remains unknown which clothing category benefits the most from the inclusive advertising strategy  \citep{joo2021impact}. Through this case study, we aim to identify the clothing category that benefits most significantly from the display of a diverse range of body shapes and investigate the performance of various experimental strategies in identifying this best-performing clothing category.

\begin{figure}[h!]
\centering \includegraphics[width=0.5\textwidth]{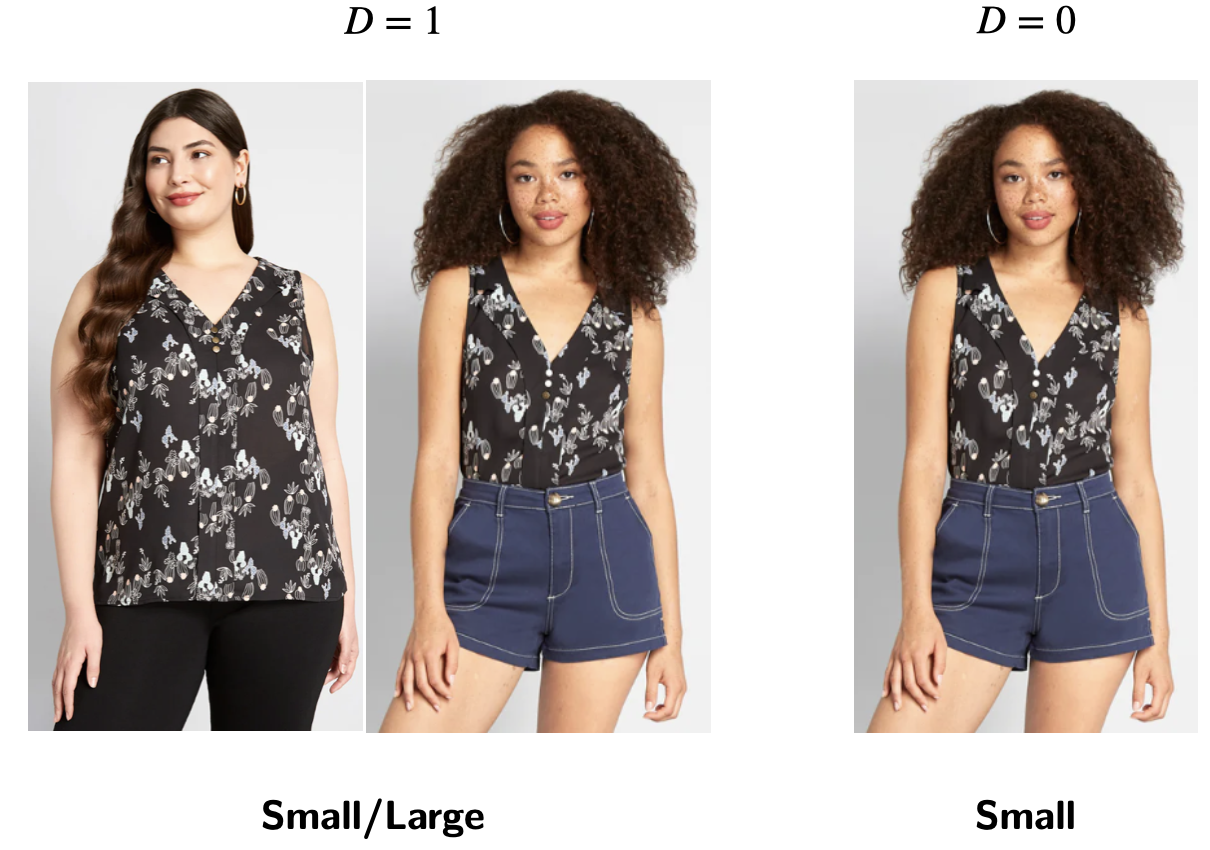}
      \caption{An example of two different advertising strategies taken from ModCloth website. The left panel shows an inclusive advertising strategy of displaying both small and plus-size human models. The right panel shows a conventional advertising strategy that only displays human models wearing size small.}
    \label{fig:modcloth}
\end{figure}

The original ModCloth data are collected and processed as in \cite{wan2020addressing}, and the dataset contains $99,893$ observations collected from 2010 to 2019. For each clothing product, the website displays one of the two types of human model images: (1) a model wearing a size small,  or (2) two models, one wearing a size small and the other a size large (Figure \ref{fig:modcloth}). We define the treatment variable as $D=1$ if both ``small" and ``large" images are displayed and $D=0$ if only ``small" images are shown. We consider four clothing categories: (1) bottoms, (2) tops, (3) outwear, and (4) dresses. In the context of our manuscript, clothing categories are equivalent to ``subgroups." To quantify customer satisfaction, we use customer ratings that range from $0$ to $5$. For this case study, we generate synthetic experimental data based on the original dataset, which shall be illustrated in the next section.

\subsection{Synthetic data generation and simulation setup}

Our data generation process mimics the ModCloth data, and we consider four non-overlapping subgroups defined by clothing categories. Denote the subgroup membership for each subject $i$ as $\bm{\mathcal{S}} = \big(\mathds{1}_{(X_i\in\mathcal{S}_1)},\ldots, \mathds{1}_{(X_i\in\mathcal{S}_4)} \big)^\intercal$. We generate the potential outcome from
$$Y_{i}(d)|X_i\in\mathcal{S}_j \sim \mathcal{N}(\mu_{d,j}, \ \sigma_{d,j}), \quad j=1, \ldots, m.$$
We obtain the subgroup mean and standard deviation parameters calibrated from the original dataset:
\begin{align*}
  \bm{\mu}_{1} &= (4.14, 4.12, 4.43, 4.48)^{\intercal}, \   \bm{\mu}_{0} = (4.83, 3.74, 4.02, 4.31)^\intercal,\\
  \bm{\sigma}_{1} &= (1.17, 1.06, 0.80, 0.90)^\intercal,\ \bm{\sigma}_{0}= (0.39, 1.57, 1.23, 1.10)^\intercal.
\end{align*}
The subgroup proportions are $\bm{p} = (0.20, 0.16, 0.56, 0.08)^\intercal$.  We denote $\bm{\tau} = (-0.69, 0.38, 0.41, 0.18)^\intercal$ as the true subgroup treatment effects. The treatment assignment $D_i$ is decided based on different experiment strategies, which shall be discussed later in the section. 
To generate synthetic data, we consider two design setups: 

\textbf{Setup 1}: We mimic the fully adaptive experiment and fix the total sample size as 
$N \in \{400,\ldots,2000\}$, and $n_1 \in \{80,\ldots, 400\}$. We assume subjects are enrolled sequentially across $T$ experimental stages, where $T\in\{320, \ldots, 1600\}$.

\textbf{Setup 2}: We mimic the multi-stage experiment and consider two settings: (a) We set $T=2$,  $n_1 \in \{300, \ldots, 1900\}$ and $n_2 = 100$. (b) We set $T=4$, $n_1 \in \{100, \ldots, 1700\}$, $n_2 = n_3 = n_4 = 100$.

\noindent Additional design setups and simulation results with a smaller first stage sample size are provided in the Supplementary Material Section E. 

Under each design setup, we compare our proposed design strategy with other conventional designs as summarized in Table \ref{table:design-strategy-simulation}. In Table \ref{table:design-strategy-simulation}
, the ``complete randomization" design refers to setting the treatment assignment probability $e_{tj} = \frac{1}{2}$ in all experimental stages, for $t = 1,\ldots, T, \ j = 1,\ldots, m$. 
The ``Neyman allocation" refers to setting the treatment assignment probability $e_{tj} = \frac{\sigma_{1j}}{\sigma_{1j}+\sigma_{0j}}$.
The ``Proposed design combined with DBCD" refers to enhancing our proposed design with the doubly adaptive biased coin design (DBCD) in \citep{zhang2006response}. The DBCD design is a response-adaptive randomization design that targets the current treatment allocation towards the optimal treatment allocation. As we consider assigning treatments to multiple subgroups, we use the DBCD design to target the optimal treatment allocation in each subgroup separately. Our implementation is summarized as follows: 
\begin{enumerate}
    \item[(1)] At Stage $t$, obtain \textit{optimal treatment allocations}  $\hat{e}_{t,j}^*$, $j=1,\ldots,m$ by solving the optimization problem as in Section \ref{subsec:adaptive-allocation-fully-adaptive}. Calculate \textit{current treatment allocation} up to Stage $t-1$, denoted as  $\hat{e}_{t-1,j}$, $j=1,\ldots,m$.
    \item[(2)] For each subgroup $j$, calculate treatment allocation under the DBCD design proposed in \cite{hu2004asymptotic}:
    \begin{align*}
        \psi_{t,j}(\hat{e}_{t,j}^*, \ \hat{e}_{t-1,j}) = \frac{\hat{e}_{t,j}^*(\frac{\hat{e}_{t,j}^*}{\hat{e}_{t-1,j}})^\gamma}{\hat{e}_{t,j}^*(\frac{\hat{e}_{t,j}^*}{\hat{e}_{t-1,j}})^\gamma + (1-\hat{e}_{t,j}^*)(\frac{1-\hat{e}_{t,j}^*}{1-\hat{e}_{t-1,j}})^\gamma},
    \end{align*}
where $\gamma \in [0,\infty)$ is a tuning parameter. 
    \item [(3)] At Stage $t$, we assign treatments with probability $\psi_{t,j}(\hat{e}_{t,j}^*, \ \hat{e}_{t-1,j})$ in each subgroup $j$.
\end{enumerate}

The ``equal enrichment design" refers to the design that sets the enrichment proportion as $p_{tj} = \frac{1}{m}$ across all the experimental stages. The ``adaptive enrichment with combination testing" approach is a method that distributes the type I error rate across experimental stages. Based on the computed type I error rate each stage aims to reach, the corresponding enrichment proportions can be estimated. We implement the combination testing approach using \texttt{R} package \texttt{rpact} \citep{lakens2021group}.

\begin{table}
\centering
\caption{Comparison of designs in our synthetic case study\label{table:design-strategy-simulation}}
\resizebox{0.9\textwidth}{!}{%
\begin{tabular}{c|c|c}
\hline\hline
 & {Fully adaptive}  & {Multi-stage}  \\
& \textbf{(Setup 1)}  &  \textbf{(Setup 2)} \\
\hline
& \multicolumn{1}{c|}{Methods in comparison}  &  Methods in comparison   \\
\cline{2-3}
\multirow{2}*{Response-adaptive design}  &   (a) Proposed design in Section \ref{subsec:adaptive-allocation-fully-adaptive}&  (a) Proposed design in Section \ref{subsec:treatment-allocation-multi-stage}  \\
& (b) Complete randomization &  \multirow{2}*{ (b) Complete randomization}   \\ 
&  {(c) Neyman allocation} &   \\
& { (d) Proposed design combined with DBCD} &   \\
\hline
\multirow{4}*{Enrichment design}  & \multirow{2}*{(a) Proposed design in Section \ref{subsec:adaptive-enrichment}}   &  (a) Proposed design in Section \ref{subsec:adaptive-enrichment} \\
 & & (b) Equal enrichment  \\
  & (b) Equal enrichment & (c) Adaptive enrichment with  \\
   & & combination testing  \\
\hline\hline
\end{tabular}}
\end{table}

We evaluate the performance of each adaptive experiment strategy from two aspects. First, we compare the experimental efforts (i.e., sample size) needed to reach various correct selection probability levels: \{$0.75,0.8, 0.85,0.9, 0.95, 0.99$\}. Second, we compare the $\sqrt{N}$-scaled bias of the estimated best subgroup treatment effect. The synthetic case study results are summarized in the following subsection.

\subsection{Synthetic case study results}

\begin{figure}[h!]
\centering
\includegraphics[width=\textwidth]{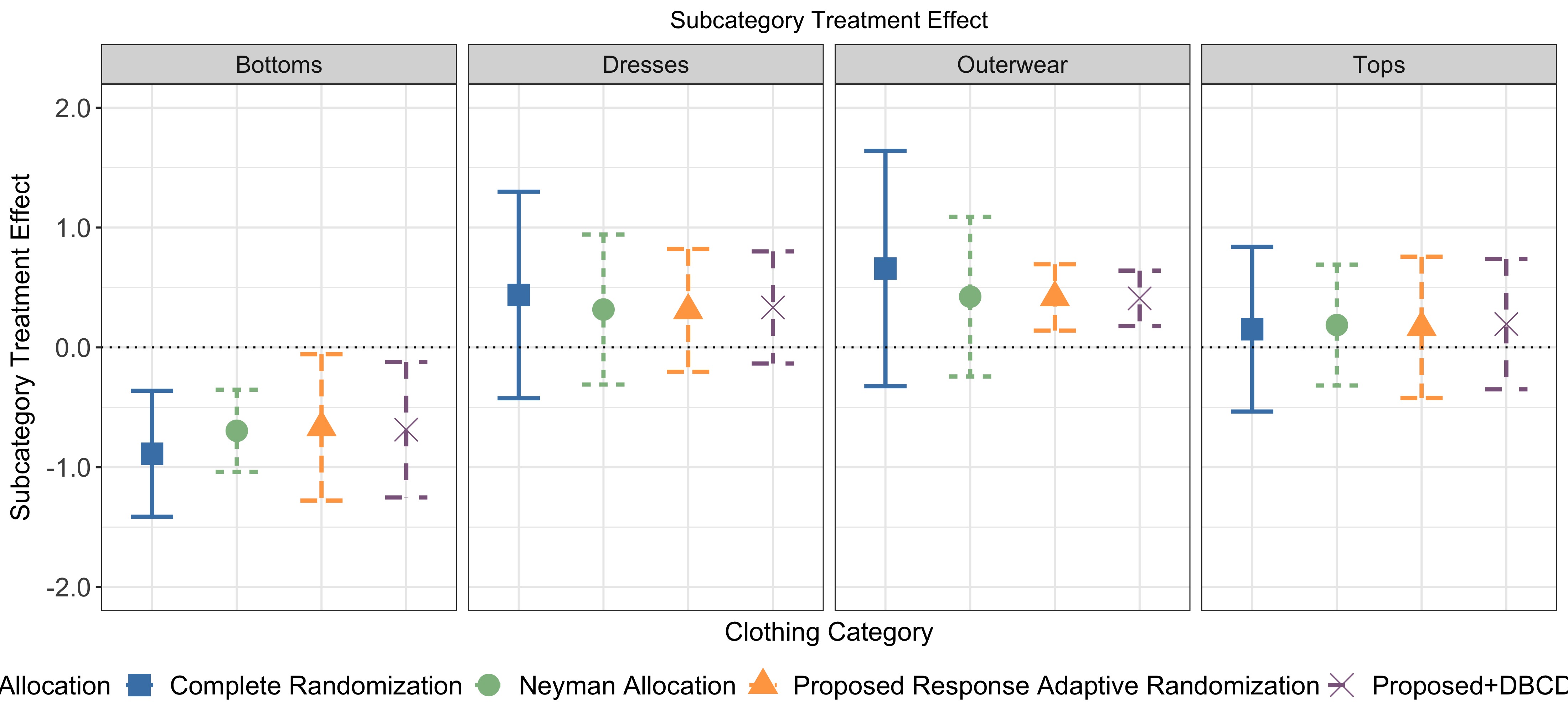}
\caption{{  The estimated treatment effects and the associated standard errors in the four clothing categories under the complete randomization design, Neyman allocation, our proposed response-adaptive randomization design, and our proposed design in combination with the DBCD design in the fully adaptive setting ($N=400$). } }
\label{fig:real-data}
\end{figure}

\begin{figure}[h!]
    \centering
    \includegraphics[width=\textwidth]{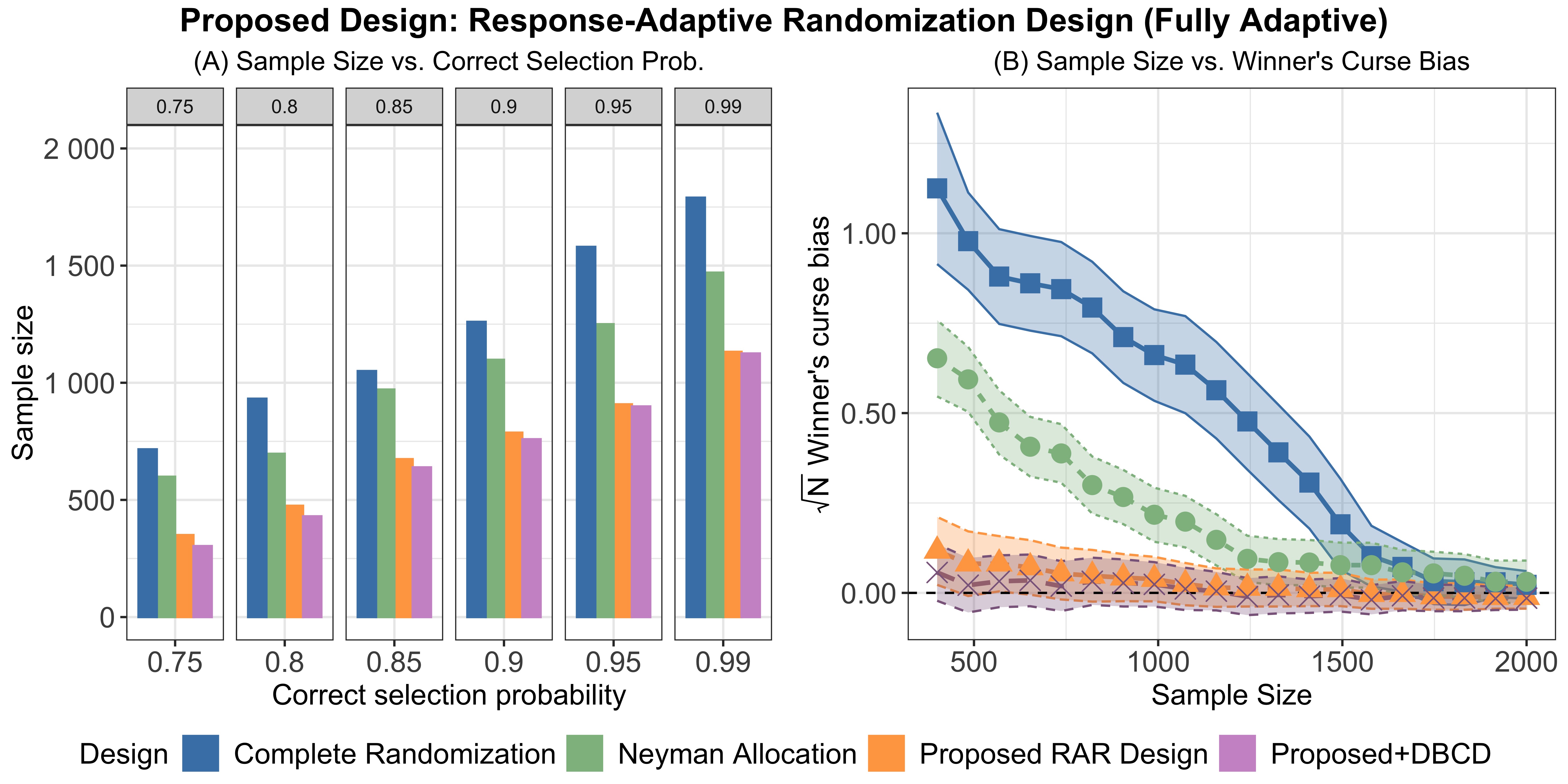}
    \caption{ {Comparison of the proposed response-adaptive randomization design, the complete randomization design, Neyman allocation, and our proposed design in combination with the doubly adaptive biased coin design under the fully adaptive setting. (A) shows the sample size comparison under various correct selection probability levels. (B) shows the $\sqrt{N}$-scaled winner's curse bias comparison with respect to different sample sizes.} }
    \label{fig:adaptive-treatment-fs}
\end{figure}

\begin{figure}[h!]
    \centering
    \includegraphics[width=0.8\textwidth]{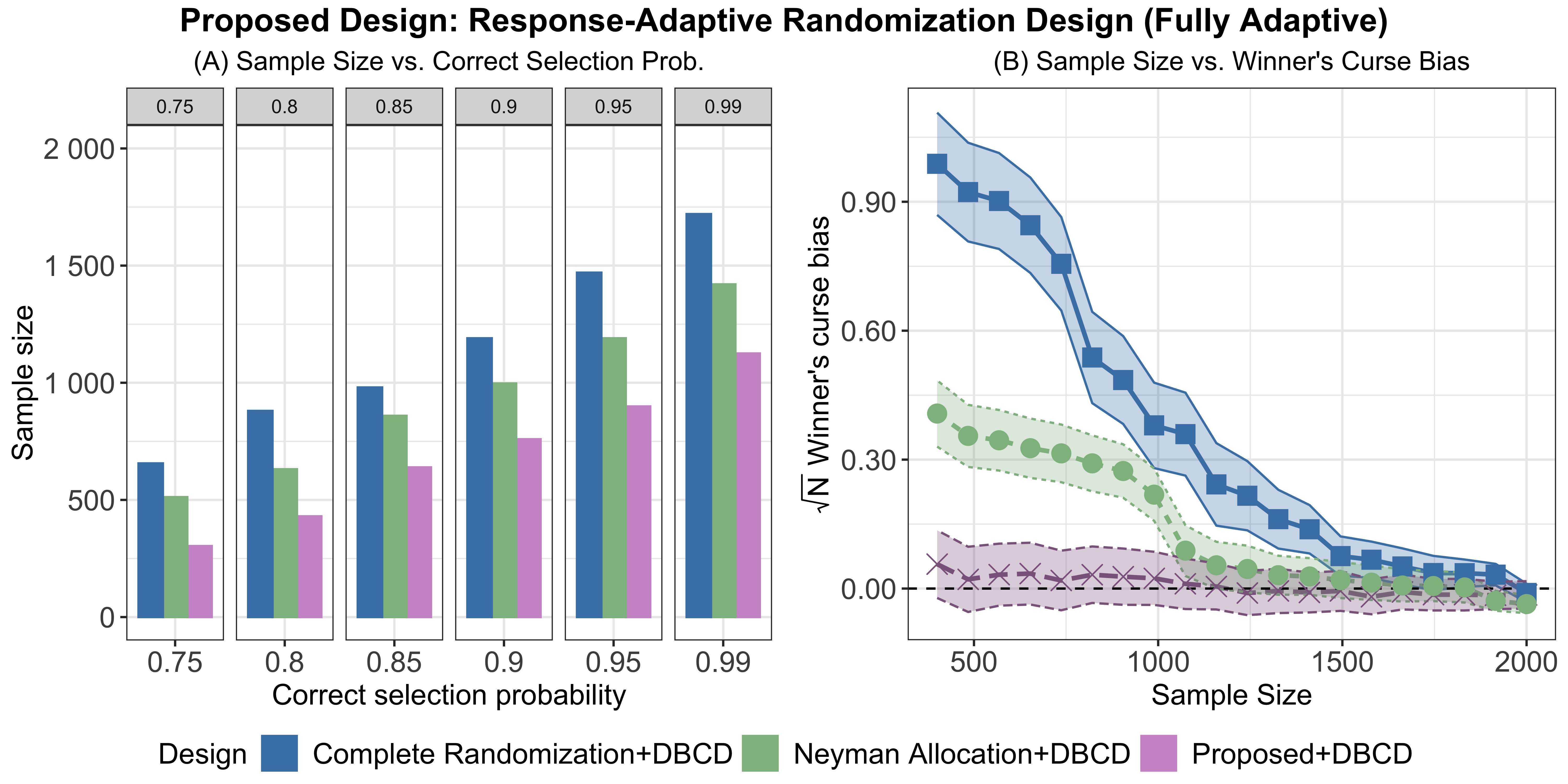}
    \caption{ { Comparison of the proposed response-adaptive randomization design, the complete randomization design, and the Neyman allocation in combination with the doubly adaptive biased coin design under the fully adaptive setting. (A) shows the sample size comparison under various correct selection probability levels. (B) shows the $\sqrt{N}$-scaled winner's curse bias comparison with respect to different sample sizes. } }
    \label{fig:adaptive-treatment-fs-dbcd}
\end{figure}

\begin{figure}[h!]
    \centering
    \includegraphics[width=\textwidth]{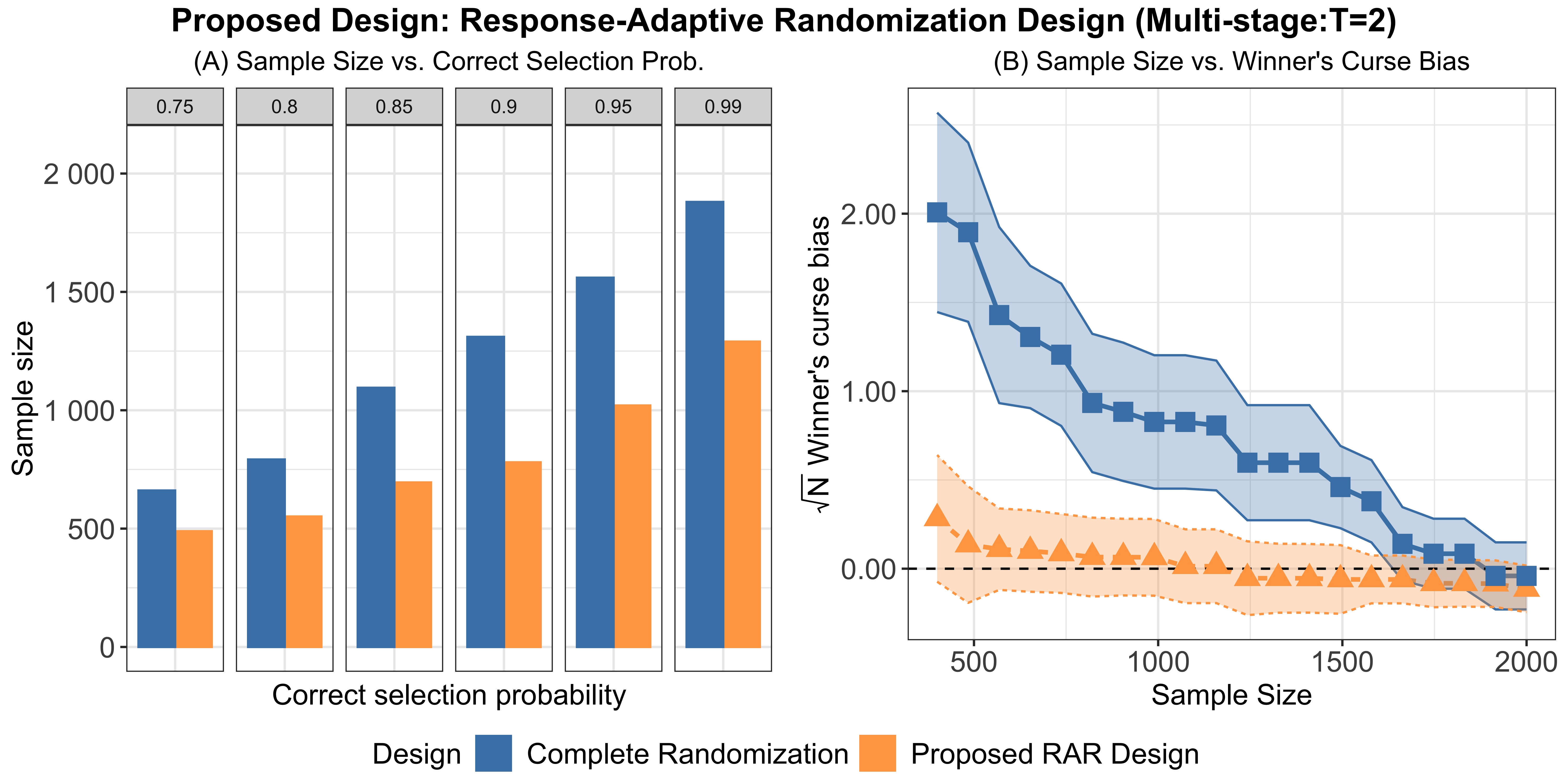}
     \includegraphics[width=\textwidth]{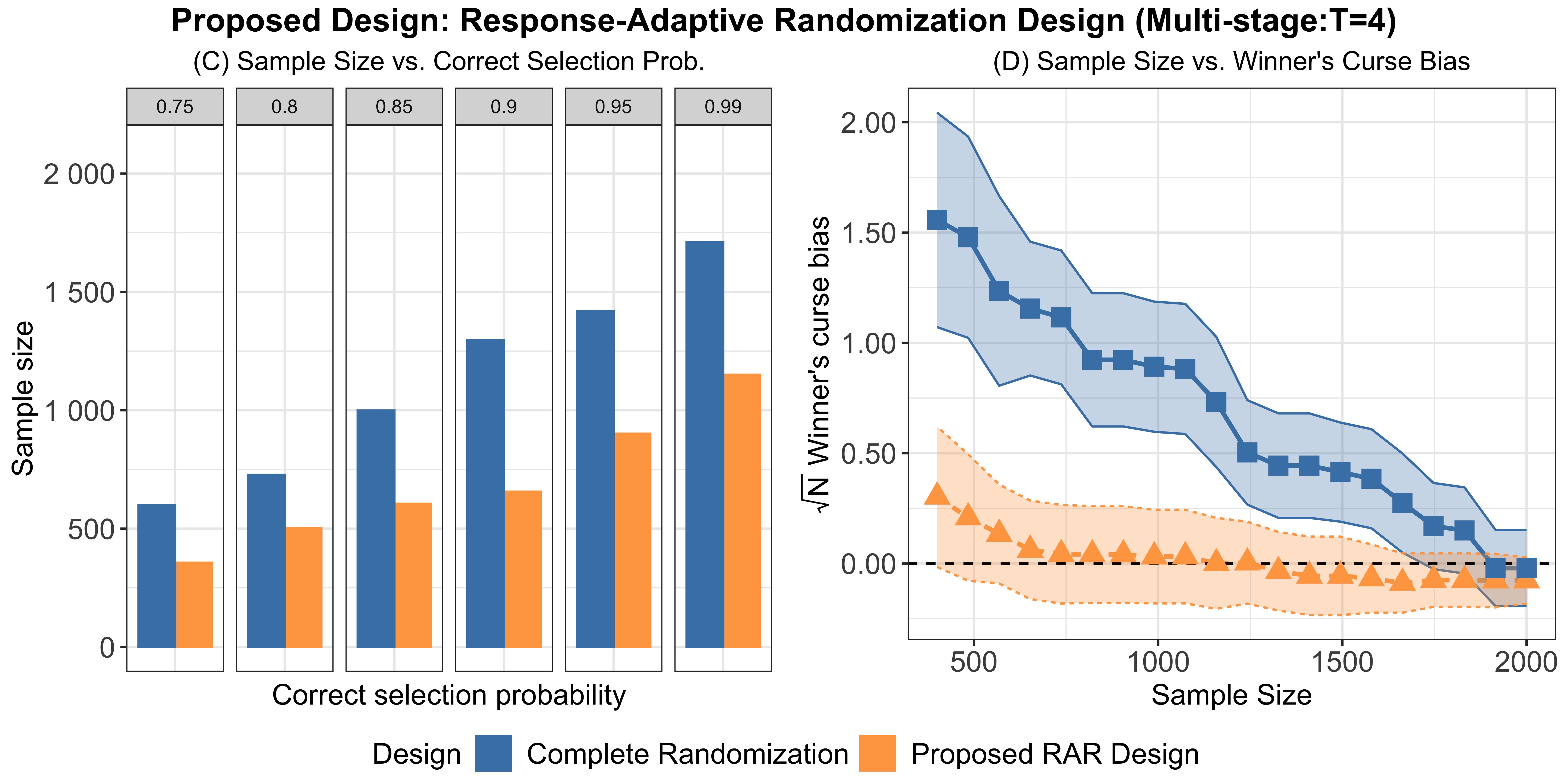}
    \caption{Comparison of the proposed response-adaptive randomization design and the complete randomization design under the multi-stage setting ($T=2$ and $T=4$). (A) shows the sample size comparison under various correct selection probability levels. (B) shows the $\sqrt{N}$-scaled winner's curse bias comparison with respect to different sample sizes.}
    \label{fig:adaptive-treatment}
\end{figure}

In Figures \ref{fig:real-data} to \ref{fig:adaptive-treatment}, we compare our proposed response-adaptive randomization design with the other conventional designs in the fully adaptive setting and the multi-stage setting. 
We summarize our simulation results from three aspects, following the order outlined at the start of Section \ref{sec:simulation}

\begin{figure}[h!]
    \centering
\includegraphics[width=\textwidth]{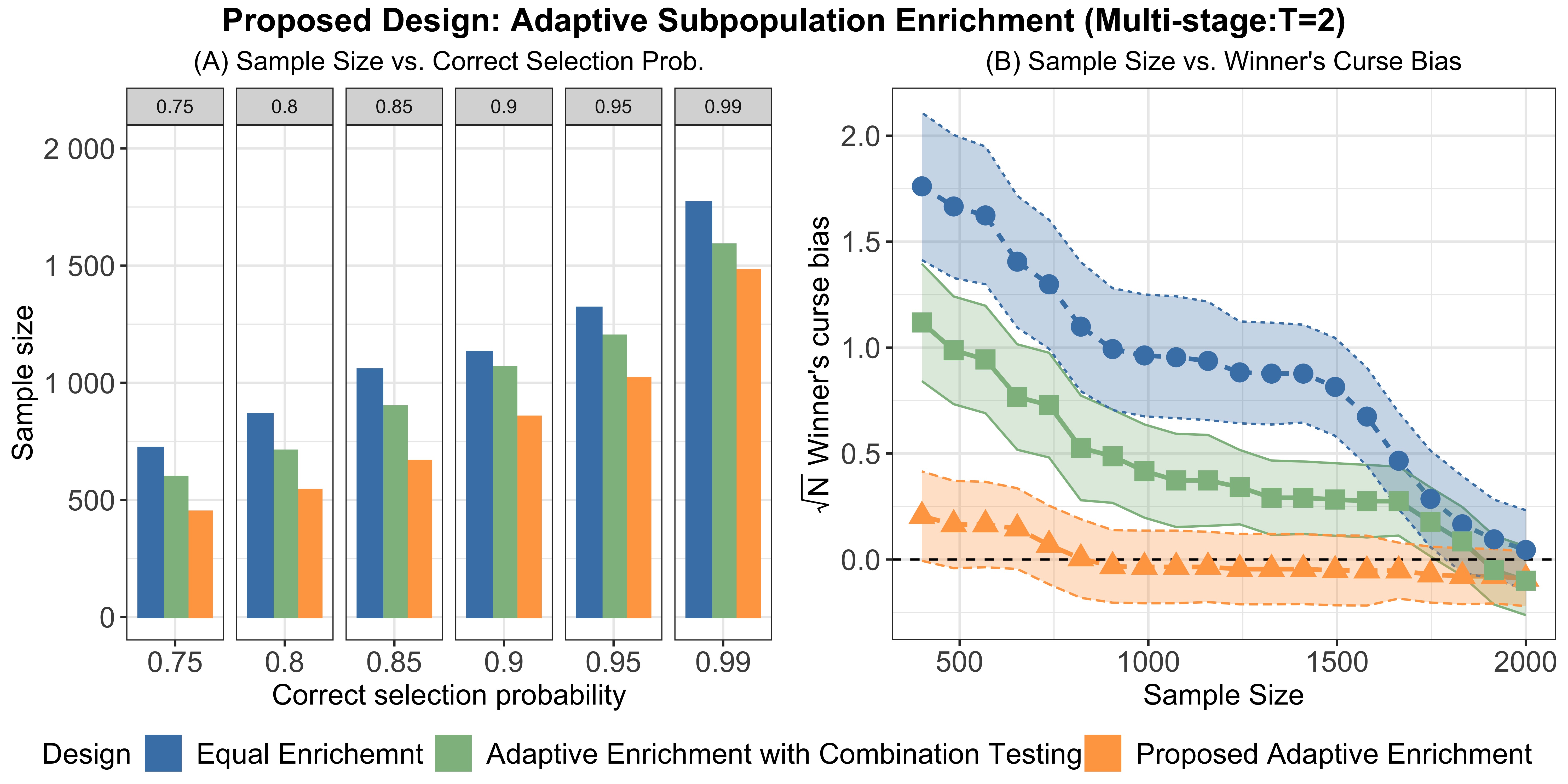}
\includegraphics[width=\textwidth]{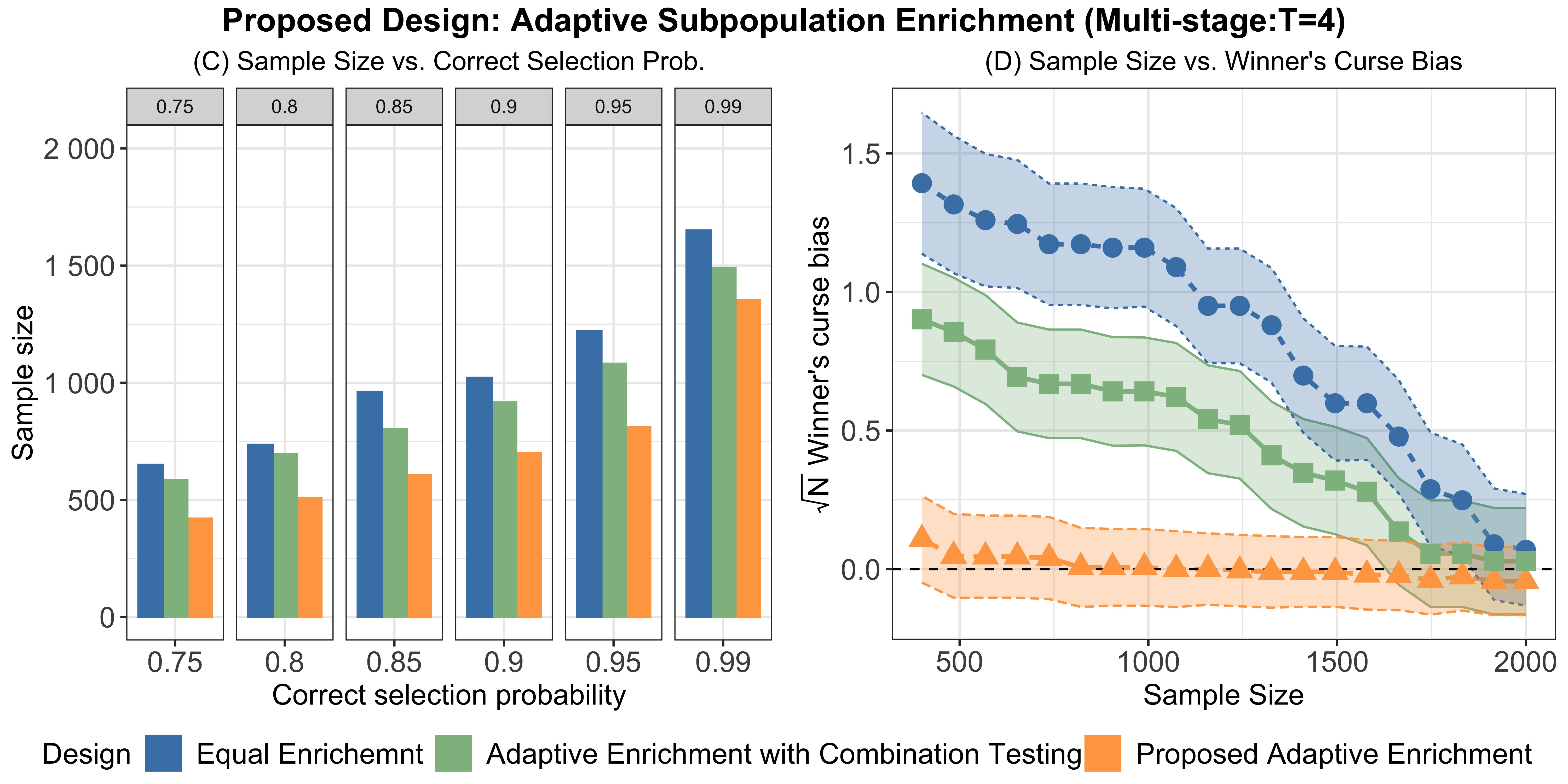}
    \caption{Comparison of the proposed adaptive subgroup enrichment design and the equal enrichment design under the multi-stage setting ($T=2$ and $T=4$). (A) shows the sample size comparison under various correct selection probability levels. (B) shows the $\sqrt{N}$-scaled winner's curse bias comparison with respect to different sample sizes.}
    \label{fig:adaptive-enrichment-ms}
\end{figure}

{\color{black} First, from the comparison in Figure \ref{fig:adaptive-treatment-fs}(A), Figures \ref{fig:adaptive-treatment} (A) and (C), and Figures \ref{fig:adaptive-enrichment-ms}(A) and (C), our proposed designs require smaller sample sizes to reach the same level of correct selection probability than other designs under comparison.} Benefiting from this design feature, our design yields the best subgroup treatment effect estimator with the lowest bias. This result supports our theoretical analysis in Proposition \ref{proposition:large-deviation-rate-comparison}.

{Second}, in line with our theoretical analysis in Proposition \ref{proposition:comparison-cr}, our proposed response-adaptive randomization design is efficient in estimating the best subgroup treatment effect and is less efficient for the worst subgroup, a trend we observe consistently in both fully adaptive and multi-stage settings. This can be seen from the results in Figure \ref{fig:real-data}. 

{Third}, the simulation results under both response-adaptive randomization design and adaptive enrichment design suggest that fully adaptive experiments can achieve equivalent levels of correct selection probability with smaller sample sizes compared to multi-stage experiments; see Figure \ref{fig:adaptive-treatment-fs}(A) and Figure \ref{fig:adaptive-treatment}(A) for example. Whenever the sample size is large, the difference between the fully adaptive and multi-stage is negligible. We conjecture that this could be attributed to the fully adaptive design providing more opportunities for experimenters to adjust treatment assignment probabilities, potentially achieving the oracle at a faster asymptotic rate.

Lastly, in the RAR setting, combining our proposed design with the DBCD design can further enhance the finite sample performance. Figures \ref{fig:real-data} 
 and \ref{fig:adaptive-treatment-fs} demonstrate that DBCD design can enhance the performance of our method in finite samples. When using the DBCD design to target our actual treatment allocation towards the optimal treatment allocation, the design strategy exhibits a smaller estimation bias for the best subgroup treatment effect and an increase in the correct selection probability. As the sample size increases, the performances of our proposed design and the DBCD-enhanced design tend to converge. In Figure \ref{fig:adaptive-treatment-fs-dbcd}, we provide an additional simulation study to highlight the broad benefits of DBCD design in enhancing the finite sample performance of various designs. We compare three designs: (i) complete randomization + DBCD, (ii) Neyman allocation + DBCD, and (iii) Proposed RAR design + DBCD. We use ``+DBCD" to indicate that the DBCD is applied to each design to guide the treatment allocation closer to the optimal treatment allocation. Figure \ref{fig:adaptive-treatment-fs-dbcd} shows that DBCD design generally improves the finite-sample performance of these design strategies by effectively targeting the current treatment allocation towards the optimal treatment allocation.

In sum, from an application perspective, when an experimenter can only enroll a limited number of subjects in online experiments, our proposed adaptive experiment strategies demonstrate more efficient use of the samples to identify the best subgroup with a higher probability and can reduce the winner's curse bias on estimating the best subgroup treatment effect. Furthermore, the results of the synthetic case study suggest that the adoption of an inclusive advertising strategy could have practical marketing advantages and potentially positive social effects. As studied in the marketing literature \citep{cinelli2016role,joo2021impact}, such an inclusive marketing strategy could improve customer satisfaction, elevate customer self-esteem, and reduce body-focused anxiety. As our designs may have applications beyond e-commons, we provide another synthetic case study in the context of health care in the Supplementary Materials Section F.

\section{Discussion}\label{sec:discussion}

In this manuscript, we propose a unified adaptive experimental framework designed to study treatment effect heterogeneity. Three directions warrant future studies. First, it is possible to extend our current framework to identify the top few subgroups instead of the best one. Take two subgroups for example; our goal can be formulated as finding the oracle treatment assignment that solves $ \max_{\bm{e}} \  \mathbb{P}\big( \min\{ \hat{\tau}_1, \hat{\tau}_2 \} \geq \max_{3\leq j\leq m} \hat{\tau}_j \big)$, 
which can be achieved by revising the objective function as $\max_{\bm{e}}\min_{3\leq j \leq m }  \max_{1\leq k \leq 2}\ G(\mathcal{S}_k, \mathcal{S}_j; e_k, e_j)$. 
Second, the theoretical analysis presented in Section \ref{subsec:efficiency-gain} is based on the assumption that the collected sample is independent and identically distributed. We hypothesize that this condition can be relaxed by utilizing refined concentration inequalities that incorporate martingales \cite{chung2006concentration, bercu2015concentration}. Exploring these possibilities will be an avenue for future research. {
Third, our design considers the setting when outcomes are observed without delay. The significance of incorporating delayed responses in adaptive trials has been discussed and recognized in various adaptive design literature, including \cite{rosenberger2012adaptive} and \cite{robertson2023response}. Some existing adaptive experiment methods and theoretical results related delayed outcomes have been discussed under the urn models \citep{wei1978randomized,wei1988exact,bai2002asymptotic,hu2004asymptotic,zhang2007generalized}, the doubly-adaptive biased coin design (DBCD) framework \citep{zhang2006response, hu2008doubly,kim2014outcome}, and in the group-sequential settings \citep{hampson2013group,schuurhuis2024two,ghosh2022robust}. We hope to extend our proposed design and adjust it for delayed outcomes in our future research. 
}

\bibliography{reference}

\end{document}